\documentclass[article,nojss,shortnames]{jss}

\usepackage[utf8]{inputenc}
\usepackage{thumbpdf,lmodern}
 
\usepackage{thumbpdf}
\usepackage{amsfonts,amstext,amsmath,amssymb}
\usepackage{accents}
\usepackage{verbatim}
\usepackage{xspace}

\usepackage[title]{appendix}

\usepackage{lineno} 

\usepackage[]{graphicx}
\usepackage[]{color}
\makeatletter
\def\maxwidth{ %
  \ifdim\Gin@nat@width>\linewidth
    \linewidth
  \else
    \Gin@nat@width
  \fi
}
\makeatother

\makeatletter
\let\ginnatheight\Gin@nat@height
\makeatother

\definecolor{fgcolor}{rgb}{0.345, 0.345, 0.345}

\usepackage{framed}
\makeatletter
 {\par\unskip\endMakeFramed%
 \at@end@of@kframe}
\makeatother

\definecolor{shadecolor}{rgb}{.97, .97, .97}
\definecolor{messagecolor}{rgb}{0, 0, 0}
\definecolor{warningcolor}{rgb}{1, 0, 1}
\definecolor{errorcolor}{rgb}{1, 0, 0}
\newenvironment{knitrout}{}{} 

\usepackage{alltt}
\IfFileExists{upquote.sty}{\usepackage{upquote}}{}
\usepackage{paralist}

\makeatletter
\g@addto@macro{\table}{\centering}
\makeatother
\usepackage{floatrow} 

\usepackage{nicefrac}

\usepackage{booktabs}

\newcommand{\rY}{Y}
\newcommand{\rX}{\mX}

\newcommand{\ry}{y}
\newcommand{\rx}{\xvec}
\newcommand{\rr}{\rvec}
\newcommand{\erx}{x}

\newcommand{\samY}{\Xi}

\newcommand{\pZ}{F} 

\newcommand{\dZ}{f}

\newcommand{\h}{h}

\newcommand{\basisy}{\avec}
\newcommand{\bern}[1]{\avec_{\text{Bs},#1}}

\newcommand{\parm}{\varthetavec}
\newcommand{\eparm}{\vartheta}
\newcommand{\dimparm}{P}

\newcommand{\shiftparm}{\betavec}
\newcommand{\scaleparm}{\gammavec}
\newcommand{\eshiftparm}{\beta}
\newcommand{\escaleparm}{\gamma}

\newcommand{\ie}{\textit{i.e.,}~}
\newcommand{\eg}{\textit{e.g.,}~}

\renewcommand{\Prob}{\mathbb{P}}
\newcommand{\Ex}{\mathbb{E}}
\newcommand{\RR}{\mathbb{R}}
\newcommand{\NN}{\mathbb{N}}

\usepackage{dsfont}

 \DeclareMathOperator{\logit}{logit}
 \DeclareMathOperator{\loglog}{loglog}
 \DeclareMathOperator{\cloglog}{cloglog}
 \DeclareMathOperator{\probit}{probit}

 \DeclareMathOperator*{\argmin}{{arg\,min}}
 \DeclareMathOperator*{\argmax}{{arg\,max}}

 \DeclareMathOperator{\ND}{N}
 
 \DeclareMathOperator{\UD}{U}

\def \avec {\text{\boldmath$a$}}

\def \mvec {\text{\boldmath$m$}}

\def \rvec {\text{\boldmath$r$}}

\def \vvec {\text{\boldmath$v$}}    
    
\def \xvec {\text{\boldmath$x$}}    \def \mX {\text{\boldmath$X$}}

 \def \calA {\mathcal A}

 \def \calI {\mathcal I}
 \def \calJ {\mathcal J}
 \def \calK {\mathcal K}
 
 \def \calM {\mathcal M}

\def \betavec         {\text{\boldmath$\beta$}}
\def \gammavec        {\text{\boldmath$\gamma$}}

\def \varthetavec     {\text{\boldmath$\vartheta$}}

\newcommand{\ubar}[1]{\underaccent{\bar}{#1}}

\usepackage{algorithm}
\usepackage{algpseudocode}
\newcommand{\alg}[1]{\normalsize\textsf{#1}}
\newcommand{\gComment}[1]{\textcolor{gray}{\Comment{#1}}}
\newcommand{\1}{\mathds{1}}
\DeclareMathOperator{\SIC}{SIC}
\DeclareMathOperator{\cor}{cor}
\newcommand{\norm}[1]{\left\lVert#1\right\rVert}

\renewcommand{\L}{Location\xspace}
\renewcommand{\l}{location\xspace}
\newcommand{\Ls}{Location-scale\xspace}
\newcommand{\ls}{location-scale\xspace}

\usepackage{xcolor}
\definecolor{Red}{rgb}{0.5,0,0}
\definecolor{Blue}{rgb}{0,0,0.5}

\title{Distribution-Free Location-Scale Regression}
\Shorttitle{Distribution-Free Location-Scale Regression}
\Plaintitle{Distribution-Free Location-Scale Regression}
 
\author{Sandra Siegfried\\Universit\"at Z\"urich \And
        Lucas Kook\\Universit\"at Z\"urich \And
        Torsten Hothorn\\Universit\"at Z\"urich}
\Plainauthor{Siegfried, Kook and Hothorn}
 
 
\Address{
Sandra Siegfried, Lucas Kook and Torsten Hothorn\\
Institut f\"ur Epidemiologie, Biostatistik und Pr\"avention \\
Universit\"at Z\"urich \\
Hirschengraben 84, CH-8001 Z\"urich, Switzerland \\
\texttt{Torsten.Hothorn@R-project.org}
}
 
\Abstract{
We introduce a generalized additive
model for location, scale, and shape (GAMLSS) next of kin aiming at 
distribution-free and parsimonious regression modelling for arbitrary
outcomes. We replace the
strict parametric distribution formulating such a model by a transformation
function, which in turn is estimated from data.  Doing so not only makes the
model distribution-free but also allows to limit the number of linear or
smooth model terms to a pair of \ls predictor functions.  We derive the likelihood
for continuous, discrete, and randomly censored observations, along with
corresponding score functions.  A plethora of existing algorithms is
leveraged for model estimation, including constrained maximum-likelihood, the
original GAMLSS algorithm, and transformation trees.  Parameter
interpretability in the resulting models is closely connected to model
selection.  We propose the application of a novel best subset selection
procedure to achieve especially simple ways of interpretation.  All
techniques are motivated and illustrated by a collection of applications
from different domains, including crossing and partial proportional hazards,
complex count regression, non-linear ordinal regression, and growth curves.
All analyses are reproducible with the
help of the \pkg{tram} add-on package to the \proglang{R} system for
statistical computing and graphics.
 }

\begin{document}

\section{Introduction}

Location-scale regression has its roots in two-sample comparisons, where one
extends the location model for some distribution function under treatment 
$F(\ry - \mu)$ by adding a scale parameter $\sigma$ to the location shift
$\mu$, that is 
$F\left({(\ry - \mu)}/{\sigma}\right)$, in comparison to the
distribution function $F(\ry)$ under no treatment. One of the earliest
contributions is Lepage's test \citep{Lepage_1971}, which is essentially a
combination of the Wilcoxon and Ansary-Bradley statistics.
Generalized additive models for location, scale,
and shape \citep[GAMLSS,][]{Rigby_Stasinopoulos_2005,GAMLSS_JSS_2007} can be
motivated as a generalization of the two-sample location-scale model to 
the regression setup, \ie with covariate-dependent location and scale
parameters, $\mu(\rx)$ and $\sigma(\rx)$, and also potentially other
parameters $\nu(\rx)$ and $\tau(\rx)$ describing skewness and kurtosis. Thus,
GAMLSS allow explanatory variables to affect multiple moments of
a variety of parametric distributions and can be understood
as an early forerunner of ``distributional'' regression models
\citep{Kneib_2021}.

For a continuous response variable $\rY$ with explanatory variables $\rX =
\rx$, GAMLSS are characterized by a parametric distribution
$\mathcal{D}$ with typically no more than four
parameters $\mu(\rx)$ for location, $\sigma(\rx)$ for scale, $\nu(\rx)$ for
skewness, and $\tau(\rx)$ for kurtosis.  For the simplest case assuming a normal
distribution $\mathcal{D} = \ND(\mu(\rx), \sigma(\rx)^2)$ for the conditional response of $\rY \in \RR$,
the model can be written 
in terms of the conditional mean $\mu(\rx)$ and standard deviation
$\sigma(\rx)$ as
\begin{eqnarray*}
\Prob(\rY \le \ry \mid \rX = \rx) = \Phi\left(\frac{\ry - \mu(\rx)}{\sigma(\rx)}\right),
\end{eqnarray*}
with $\Phi$ being the standard normal cumulative distribution function.
Without relying on such a prior assumption of a parametric distribution $\mathcal{D}$,
\citet[Equation~1]{Tosteson_1988} introduced a distribution-free
\ls ordinal regression model in the context of receiver operating characteristic (ROC)
analysis for ordinal responses $\rY \in \{\ry_1 < \ry_2 < \cdots < \ry_K\}$
formulated as a conditional cumulative distribution function
\begin{eqnarray} \label{eq:TB}
\Prob(\rY \le \ry_k \mid \rX = \rx) =
\pZ\left(\frac{\eparm_k - \mu(\rx)}{\sigma(\rx)}\right), \quad k = 1, \dots, K - 1
\end{eqnarray}
with intercept thresholds $\eparm_k$ depending on the $k$th response category, parameters
$\eparm_k \leq \eparm_{k + 1}$ being monotonically non-decreasing. 
The model features two model terms, $\mu(\rx)$ and $\sigma(\rx)$, and
is defined by a cumulative distribution function $\pZ$. \cite{Tosteson_1988}
discuss normal ($\pZ = \Phi$) and logit models ($\pZ = \logit^{-1}$) in more
detail.
The latter corresponds to the ``non-linear'' odds model discussed in
\citet[Section~6.1]{McCullagh_1980}, which was later extended in terms of
``partial proportional odds models'' \citep{Peterson_Harrell_1990}. A very
attractive feature of such models is their distribution-free nature and
easily comprehensible covariate-dependence through location-scale parameters
$\mu(\rx)$ and $\sigma(\rx)$. However, they lack the broad applicability of
the GAMLSS family, for example to censored, bounded or mixed
discrete-continuous responses. Inspired by 
\ls ordinal regression our primary aim is to develop a
distribution-free and parsimonious flavor of GAMLSS.

We propose a generalization of \ls ordinal regression by introducing a smooth
parsimonious parameterization of the intercept thresholds in terms of a
transformation function, allowing to estimate distribution-free \ls models for
continuous, discrete, and potentially censored or truncated outcomes in a unified
maximum likelihood framework \citep{Hothorn_Moest_Buehlmann_2017}.  This
framework of \ls transformation models allows one to model the impact
of explanatory variables on the location and the dispersion of the
response distribution, without relying on distributional assumptions. We demonstrate
the practical merits of such an approach by applications of
(1) maximum-likelihood estimation in stratified models and models for crossing or
partially proportional hazards, (2) novel location-scale regression trees,
(3) transformation models with smooth non-linear location-scale parameters
for growth-curve analysis, and discuss (4) model selection issues 
arising in these contexts.

\section{Model} \label{sec:model}

\begin{figure}[t!]
\begin{knitrout}\small
\definecolor{shadecolor}{rgb}{1, 1, 1}\color{fgcolor}

{\centering \includegraphics{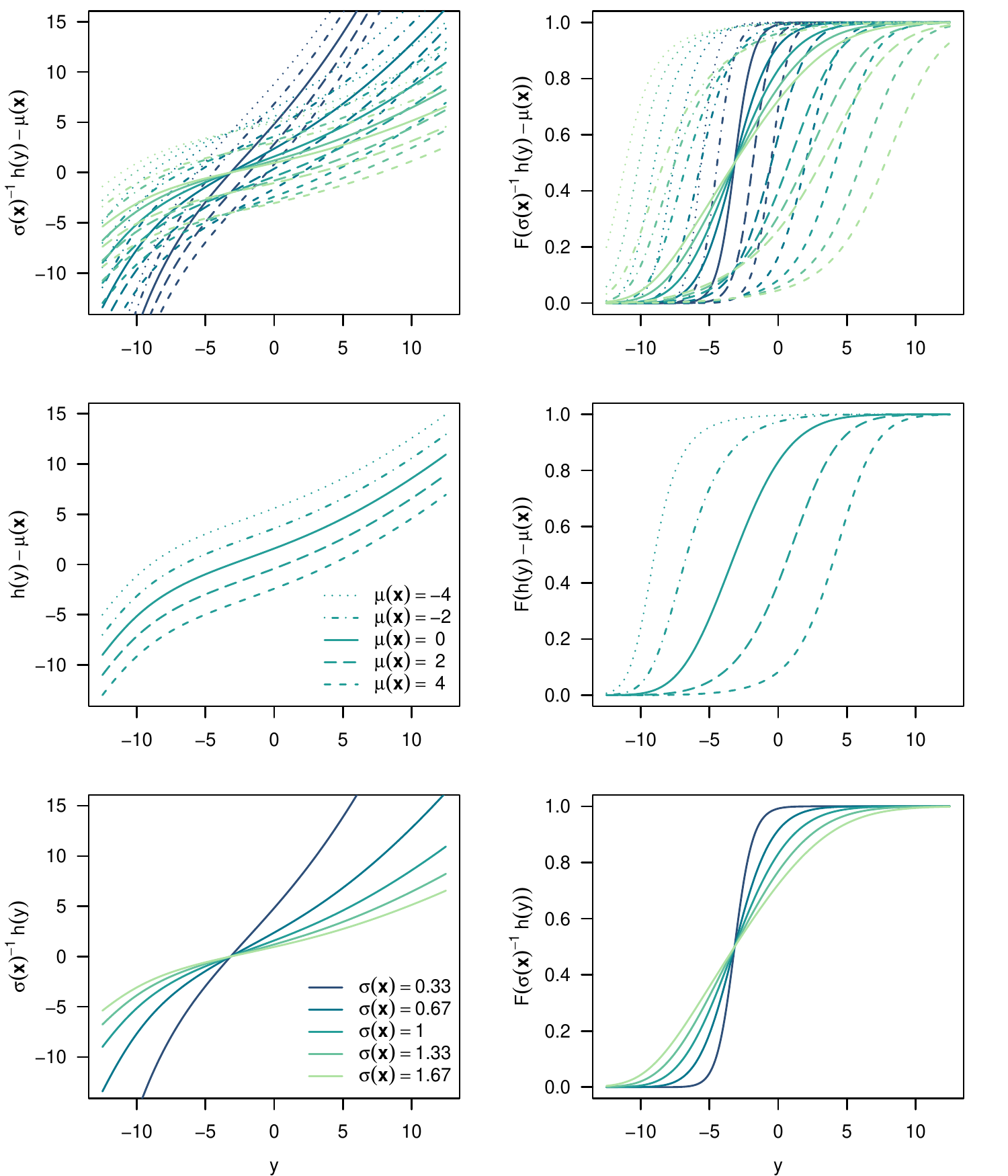} 

}

\end{knitrout}
\caption{
\Ls transformation model.  The transformation (left) and cumulative
distribution function (right) are shown for the baseline configuration (\ie~$\mu(\rx) =
0$ and $\sigma(\rx) = 1$) and different values of the location parameter
$\mu(\rx)$ and of the scale parameter $\sigma(\rx)$.
\label{fig:stram}}
\end{figure}

For univariate and at least ordered responses variables $\rY \in \samY$ we
propose to study regression models describing the conditional distribution
function of $\rY$ given explanatory variables $\rX = \rx$ as
\begin{eqnarray}
\Prob(\rY \le \ry \mid \rX = \rx) =
  \pZ\left(\sigma(\rx)^{-1} \h(\ry \mid \parm) - \mu(\rx)\right), \quad \ry \in \samY.
\label{eq:model}
\end{eqnarray}
The model is characterized by (i) a monotonically increasing transformation
function $\h: \samY \rightarrow \RR$ depending on parameters $\parm \in
\RR^\dimparm$, (ii) a cumulative distribution function $\pZ: \RR \rightarrow
[0, 1]$ of some random
variable with log-concave Lebesgue density on the real line, (iii) a
covariate-dependent location parameter $\mu(\rx) \in \RR$, and (iv) a
covariate-dependent scale parameter $\sigma(\rx) \in \RR^+$.  The model is
distribution-free in the sense that a unique transformation function $\h$
exists for every baseline distribution $\Prob(\rY \le \ry \mid \rX =
\rx_0)$, \ie a distribution conditional on explanatory variables $\rx_0$
with $\mu(\rx_0) = 0$ and $\sigma(\rx_0) = 1$.  In this case, the
transformation function is given by $\h(\ry \mid \parm) = \pZ^{-1}(\Prob(\rY
\le \ry \mid \rX = \rx_0))$.  Conditional distributions arising from
changing $\rx_0$ to $\rx$ are linear in $\h$ on the scale of the link
function $\pZ^{-1}(\Prob(\rY \le \ry \mid \rX = \rx)) = \sigma(\rx)^{-1}
\h(\ry \mid \parm) - \mu(\rx)$. Unknowns to be estimated are the
parameters $\parm$ defining the transformation function, the location
function $\mu(\rx)$, and the scale function $\sigma(\rx)$, whereas $\pZ$ is
chosen \emph{a priori}. 
The applicability to ordered, count, or continuous
outcomes possibly under random censoring, its distribution-free nature, and the
\ls formulation allowing simple interpretation of the impact explanatory
variables have on the response' distribution shall be discussed in the
following.
Figure~\ref{fig:stram} illustrates the flexibility of the \ls transformation model on the scale
of the link function, \ie~$\pZ^{-1}(\Prob(\rY \le \ry \mid \rX = \rx))$, and
of the conditional distribution function $\Prob(\rY \le \ry \mid \rX = \rx)$ for different values of $\mu(\rx)$ and $\sigma(\rx)$.

\subsection{Interpretation} \label{sec:inter}

In all models~(\ref{eq:model}), positive values of the \l parameter
$\mu(\rx)$ correspond to larger values of the response and smaller values of the
scale parameter $\sigma(\rx)$ are associated with smaller variability of
the response and thus result in more ``concentrated'' conditional
distributions (Figure~\ref{fig:stram}). 
Fitted models can conveniently be inspected on the scale of the
conditional distribution, survival, density, (cumulative) hazard, odds, or quantile
functions.

Statements beyond these general facts and interpretation of
$\mu(\rx)$ and $\sigma(\rx)$ in particular depend on the specific choice of
$\pZ$.  Suitable choices for $\pZ$ include inverses of common link
functions, such as $\pZ = \Phi = \probit^{-1}$, $\pZ = \logit^{-1}$, $\pZ =
\cloglog^{-1}$, or $\pZ = \loglog^{-1}$.  For $\sigma(\rx) \equiv 1$, the
model reduces to well-established regression models.  For $\pZ =
\cloglog^{-1}$, one obtains a proportional hazards model, a proportional
reverse-time hazards model is defined by $\pZ = \loglog^{-1}$, and a
proportional odds model given by the choice $\pZ = \logit^{-1}$.  For $\pZ =
\Phi$, $\mu(\rx) = \Ex(\h(\rY \mid \parm) \mid \rx)$ is the conditional mean
of the $\h$-transformed response. An overview on these models and
interpretation of $\mu(\rx)$ is available from
\citet[][Table~1]{Hothorn_Moest_Buehlmann_2017}.

Under certain circumstances, these simple ways of interpretation carry
over to \ls models of form~(\ref{eq:model}).  Consider Cox' proportional hazards
model ($\pZ = \cloglog^{-1}$) for a continuous survival time.  A change from
$\rx$ to $\tilde{\rx}$ is reflected by the difference $\mu(\tilde{\rx}) -
\mu(\rx)$ on the scale of the log-hazard functions conditional on $\rx$ and
$\tilde{\rx}$, respectively.  The introduction of a scale parameter
$\sigma(\rx)$ to this model does not affect this form of interpretation as
long as $\sigma(\rx) = \sigma(\tilde{\rx})$, owing to the fact that in
model~(\ref{eq:model}) $\mu(\rx)$ is not multiplied with $\sigma(\rx)^{-1}$.
For a proportional odds model,
$\mu(\tilde{\rx}) - \mu(\rx)$ is the vertical difference between the two
conditional log-odds functions. Therefore, if model interpretation on
these scales is important for certain explanatory variables, one should try
to omit these variables from the scale term.

Another form of model interpretation can be motivated from 
probabilistic index models \citep{Thas_Neve_Clement_2012}, which
describe the impact of a transition from $\rx$ to $\tilde{\rx}$
by the probabilistic index $\Prob(\rY \leq \tilde\rY \mid \rx,
\tilde\rx)$.
This probability can be derived from transformation
models \citep{Sewak_Hothorn_2022}. For the probit
\ls transformation model $\Phi(\sigma(\rx)^{-1} \h(\ry \mid
\parm) - \mu(\rx))$, for example, the probabilistic index has a simple form,
%
%
\begin{eqnarray*}
\Prob\left(\rY \leq \tilde\rY \mid \rx, \tilde\rx\right) =
\Prob\left(\h(\rY \mid \parm) \leq \h(\tilde\rY \mid \parm) \mid \rx, \tilde\rx\right) =
\Phi\left(\frac{\sigma(\tilde{\rx})\mu(\tilde{\rx}) -
\sigma(\rx)\mu(\rx)}{\sqrt{\sigma(\tilde{\rx})^2 + \sigma(\rx)^2}}\right),
\end{eqnarray*}
with two independent draws from this model, the first, $\rY$, conditional on
$\rx$ and the second, $\tilde{\rY}$,
conditional on $\tilde{\rx}$.
%
%
Especially in cases where an explanatory variable affects both the \l term
$\mu(\rx)$ but also the scale term $\sigma(\rx)$, the probabilistic index may serve as a
comprehensive measure to describe the impact of changes in the covariate
configuration on the response's distribution.

\subsection{Parameterization}

We in general express the transformation function in terms of $\dimparm$
basis functions $\h(\ry \mid \parm) = \basisy(\ry)^\top \parm$.
For absolute continuous responses $\rY \in \samY \subseteq \RR$, 
the transformation function $\h$ can be conveniently parameterized in terms
of a polynomial in Bernstein form $\h(\ry \mid \parm) = \bern{\dimparm - 1}(\ry)^\top \parm$ 
\citep{McLain_Ghosh_2013,Hothorn_Moest_Buehlmann_2017}. The
basis functions $\bern{\dimparm - 1}(\ry) \in \RR^\dimparm$ are specific beta densities
\citep{Farouki_2012} and it is straightforward to
obtain derivatives and integrals of $\h(\ry \mid \parm) = \bern{\dimparm -
1}(\ry)^\top \parm$ with respect to $\ry$ and, under suitable constraints, a
monotonically increasing transformation function $\h$
\citep{Hothorn_Moest_Buehlmann_2017}.
For count responses $\rY \in \{0, 1, 2, \dots\}$ this transformation
function is evaluated for integer values only,
\ie~$\h(\lfloor\ry\rfloor \mid \parm)$ \citep{Siegfried_Hothorn_2020}.
For ordered categorical responses $\rY \in \{\ry_1 < \cdots < \ry_K\}$ the
transformation function is defined such that $\basisy(\ry_k)^\top \parm = \eparm_k$ 
depending on the category $k = 1, \dots, K - 1$. A non-parametric version assigns one
parameter to each unique value of the outcome in the same way. In all cases,
monotonicity of $\h$ can be implemented by the constraints
$\eparm_p \le \eparm_{p + 1}, p \in 1, \dots, \dimparm - 1$ \citep{Hothorn_Moest_Buehlmann_2017}.

\subsection{Likelihood}

From model~(\ref{eq:model}), the log-likelihood contribution $\ell_i(\parm, \mu(\rx_i),
\sigma(\rx_i))$ of an
observation $(\ry_i, \rx_i)$ with $\ry_i \in \RR$ given as a function of
the unknown parameters $\parm$, $\mu(\rx_i)$, and $\sigma(\rx_i)$ is
\begin{eqnarray} \label{eq:cll}
\log\left[\dZ\left\{\sigma(\rx_i)^{-1} \h(\ry_i \mid \parm) - \mu(\rx_i) \right\}\right] +
    \log\left[\sigma(\rx_i)^{-1}\right] + \log\left[\h^\prime(\ry_i \mid \parm)\right].
\end{eqnarray}
Evaluating this expression requires the Lebesgue density $\dZ = \pZ^\prime$ and
the derivative $\h^\prime(\ry \mid \parm ) = \basisy^\prime(\ry)^\top \parm$
of the transformation function with respect to $\ry$.
For a discrete, left-, right- or interval-censored observation $(\ubar{\ry}_i, \bar{\ry}_i]$ 
the exact log-likelihood contribution $\ell_i(\parm, \mu(\rx_i),
\sigma(\rx_i)) = \log\{\Prob(\rY \in (\ubar{\ry}_i, \bar{\ry}_i] \mid \rx_i)\}$ is
\begin{eqnarray} \label{eq:dll}
\log\left[\pZ\left\{\sigma(\rx_i)^{-1} \h(\bar{\ry}_i \mid \parm) -  \mu(\rx_i)\right\} -
     \pZ\left\{\sigma(\rx_i)^{-1} \h(\ubar{\ry}_i \mid \parm) - \mu(\rx_i)\right\}\right].
\end{eqnarray}
For observed categories $\ry_k$, the datum is specified by $(\ry_{k - 1}, \ry_k]$
and for counts $\ry_i \in \NN$ it is $(\ubar{\ry}_i, \bar{\ry}_i] = (\ry_i - 1, \ry_i]$.
For random right-censoring at time $t_i$ it is given by $(\ubar{\ry}_i, \bar{\ry}_i] = (t_i, \infty)$
and for left-censoring at time $t_i$ by $(\ubar{\ry}_i, \bar{\ry}_i] = (0, t_i]$.

For the important special case of $i = 1, \dots, N$ independent realizations
from model~(\ref{eq:model}) with linear \l term $\mu(\rx) = \rx^\top\shiftparm$ and 
log-linear form for the
scale term $\sigma(\rx)^{-1} = \sqrt{\exp(\rx^\top\scaleparm)}$, the unknown parameters
$\parm$, $\shiftparm$, and $\scaleparm$ can be estimated simultaneously 
by maximizing the corresponding log-likelihood under suitable constraints
\begin{eqnarray*}
   (\hat{\parm}_N, \hat{\shiftparm}_N, \hat{\scaleparm}_N) =
   \argmax_{\parm, \shiftparm, \scaleparm}
   \sum_{i = 1}^N \ell_i\left(\parm, \rx_i^\top\shiftparm,
   \sqrt{\exp(\rx_i^\top \scaleparm)}^{-1}\right) \\
   \text{subject to } \eparm_p \le \eparm_{p + 1}, p \in 1, \dots, \dimparm - 1.
\end{eqnarray*}
Score functions and Hessians as well as conditions for likelihood-based
inference can be derived from the expressions given in \cite{Hothorn_Moest_Buehlmann_2017}
for models defined in terms of $\parm$ and $\shiftparm$.

\newpage
\subsection{Model selection} \label{sec:modsel}

Model selection in this framework can
be performed by including an $L_0$~penalty in the likelihood implied by 
model~(\ref{eq:model})
\begin{align*}
  \max_{\parm \in \RR^\dimparm, \shiftparm \in \RR^J, \scaleparm \in \RR^J} 
  \sum_{i=1}^N \ell_i\left(\parm, \rx_i^\top\shiftparm, 
  \sqrt{\exp(\rx_i^\top\scaleparm)}^{-1}\right), \quad \mbox{subject to } 
  \norm{\left(\shiftparm^\top, \scaleparm^\top\right)^\top}_0 \leq s,
\end{align*}
using an adaptation of the sequencing-and-splicing technique suggested by 
\cite{Zhu_2020}. Here, $s \in \{1, \dots, 2J\}$ denotes a fixed support size
and $\norm{\boldsymbol\cdot}_0$ denotes the $L_0$~norm. The parameters of the transformation 
function $\parm$ remain unpenalized. When the support size $s$ is unknown, $s$ is 
tuned by minimizing a high-dimensional information criterion (SIC). The procedure is 
summarized in Algorithm~\ref{alg:sabess}. Further information on the choice of
the initial active set and the inclusion of unpenalized parameters is given in 
the Supplementary Material~\ref{sec:bss}.
\begingroup 
\renewcommand{\tablename}{Algorithm}
\begin{algorithm}[!ht]
\footnotesize
\setlength{\abovedisplayskip}{.2pt}
\setlength{\belowdisplayskip}{.2pt}
\setlength{\abovedisplayshortskip}{.2pt}
\setlength{\belowdisplayshortskip}{.2pt}
\caption{Best subset selection for \ls transformation models.}\label{alg:sabess}
\begin{algorithmic}[1]
\Require Data $\{(\ry_i, \rx_i)\}_{i=1}^N \in (\Xi \times \RR^J)^N$, 
    max. support size $s_\text{max} \in \{1, \dots, 2J\}$, 
    max. splicing size $k_\text{max} \leq s_\text{max}$, tuning threshold 
    $\tau_s \in \RR_+$ for $s = 1, \dots, s_\text{max}$
\State Fit unconditional model: $\quad \hat\parm \gets \argmax_{\parm \in \RR^P} \sum_{i=1}^N 
        \ell_i(\parm, 0, 1)$
\State Compute bivariate score residuals:
\begin{align*}
  (r_\text{loc}, r_\text{sc})_i \gets \left.
    \frac{\partial}{\partial(\eshiftparm, \escaleparm)} 
        \ell_i\left(\parm, \eshiftparm, \sqrt{\exp(\escaleparm)}^{-1}\right)\right\rvert_{\parm=\hat\parm,\eshiftparm=0, \escaleparm=0}
\end{align*}
\For{$s = 1, 2, \dots, s_\text{max}$}
\State Initialize active set: 
\begin{align*}
\calA^0_s &= \bigg\{i : 
\sum_{j = 1}^J
    \1\left(
        \lvert\cor(\rx_j, \rvec_\text{loc})\rvert \geq \lvert\cor(\rx_i, \rvec_\text{loc})\rvert 
    \right) \; +\\
&\sum_{k = J + 1}^{2J}
    \1\left(
        \lvert\cor(\rx_k, \rvec_\text{sc})\rvert \geq \lvert\cor(\rx_i, \rvec_\text{sc})\rvert 
    \right)
    \leq s \bigg\}, \\
    \calI^0_s &= \{1, \dots, 2J\} \backslash \calA^0_s
\end{align*} 
    \For{$m = 0, 1, 2, \dots$} 
        \State Run Algorithm~2 in \cite{Zhu_2020}:
        \begin{align*}
        \left(\hat\parm^{m+1}_s, \hat\shiftparm^{m + 1}_s, \hat\scaleparm^{m+1}_s, \calA^{m+1}_s, 
        \calI^{m+1}_s\right)
        \gets\\ \operatorname{Splicing}(\hat\parm^m_s, \hat\shiftparm^m_s, \hat\scaleparm^m_s, \calA^m_s, 
        \calI^m_s, k_\text{max}, \tau_s)
        \end{align*}
    \If{$\left(\calA^{m+1}_s, \calI^{m+1}_s\right) = \left(\calA^m_s, \calI^m_s\right)$} 
    \State stop
    \EndIf
    \EndFor
\State $\left(\hat\parm_s, \hat\shiftparm_s, \hat\scaleparm_s, \hat\calA_s\right) \gets 
    \left(\hat\parm^{m+1}_s, \hat\shiftparm^{m+1}_s, \hat\scaleparm^{m+1}_s, \calA^{m+1}_s\right)$
\EndFor
\State Choose optimal support size based on SIC:
  \begin{align*}
    \hat s = 
\argmin_{s} - \sum_{i=1}^N \ell_i\left(\hat\parm_s, \rx_i^\top\hat\shiftparm_s, 
    \sqrt{\exp(\rx_i^\top\hat\scaleparm_s)}^{-1}\right) \ + \\
    \norm{\left(\hat\shiftparm_s^\top, \hat\scaleparm_s^\top\right)^\top}_0
    (\log 2J) (\log\log N)
  \end{align*}
\State \Return{$(\hat\parm_{\hat s}, \hat\shiftparm_{\hat s}, \hat\scaleparm_{\hat s}, \hat \calA_{\hat s})$}
\end{algorithmic}
\end{algorithm}
\endgroup
\makeatletter
\ifcsname c@posttable\endcsname
  \addtocounter{posttable}{-1}
\else 
\fi
\makeatother
 
\newpage
\section{Inference for applications}\label{sec:estimation}

Motivated by applications from different domains we detail 
the estimation of \ls transformation models,
including important aspects of model evaluation, interpretation and testing.
The wide range of applications of the model framework is further exemplified by
contrasting it to established \ls models. 

In Section~\ref{sec:STRAT} we outline the estimation of \ls transformation models
from the perspective of a stratified model. Section~\ref{sec:XH} presents
the application of \ls models to survival data in the presence of crossing hazards, 
further introducing a \ls alternative to the commonly used log-rank test.
Interpretability of \ls transformation models is exemplified in Section~\ref{sec:PPH}
assessing seasonal and annual patterns of deer-vehicle collisions. The estimation of
non-linear, tree-based \ls transformation models is discussed for 
self-reported orgasm frequencies of Chinese women in Section~\ref{sec:TRTF}.
Inspired by the GAMLSS framework, Section~\ref{sec:TAMLS} describes the estimation 
of a distribution-free version of additive models featuring smooth covariate-dependent
\l and scale terms.
The application of model selection in this framework
is exemplified in Section~\ref{sec:VS}.

All analyses can be replicated using the \pkg{tram} \proglang{R}~package \citep{pkg:tram} with
\begin{verbatim}
R> demo("stram", package = "tram")
\end{verbatim}
%

\subsection{Maximum likelihood} \label{sec:ML}

\subsubsection{Stratification}\label{sec:STRAT}
\cite{Haslinger_Korte_Hothorn_2020} reported data from measuring postpartum blood
loss $\rY \in \RR^+$ during 676~vaginal deliveries
and 632~caesarean sections
at the University Hospital Zurich, Switzerland. Aiming to contrast 
blood loss during vaginal deliveries or cesarean sections the conditional distributions
can be estimated by stratification, for example.

In the following we estimate such a stratified model with two separate transformation functions
$\h(\ry \mid \text{delivery} = \text{vaginal}) = \bern{15}(\ry)^\top
\parm_\text{vaginal}$ and $\h(\ry \mid \text{delivery} = \text{cesarean}) =
\bern{15}(\ry)^\top \parm_\text{cesarean}$ as polynomials in Bernstein form.
In a similar spirit, a \ls transformation model with transformation
function $\bern{15}(\ry)^\top \parm$ for vaginal
deliveries and transformation function $\sqrt{\exp(\escaleparm)}
\bern{15}(\ry)^\top \parm - \eshiftparm$ for cesarean sections can be defined. Transformation functions of both models
were defined on the probit scale. The stratified, distribution- and model-free
approach estimates $2 \times \dimparm = 32$ parameters
whereas the distribution-free \ls model consists of only $\dimparm + 2 =
18$ parameters, providing a lower-dimensional 
alternative to stratification. Owing to Weierstrass' theorem, polynomials in Bernstein
form with sufficiently large order $\dimparm - 1$ can approximate any continuous function
on an interval and therefore the \ls model does not make assumptions
about the transformation function $\h$ and thus the distribution of blood loss
for vaginal deliveries. However, the location and scale terms govern the
discrepancy between blood loss distributions for the two modes of delivery,
and thus this approach can be characterized as being distribution-free but not model-free.

\begin{figure}[t!]
\centering
\begin{knitrout}\small
\definecolor{shadecolor}{rgb}{1, 1, 1}\color{fgcolor}

{\centering \includegraphics[width=\textwidth]{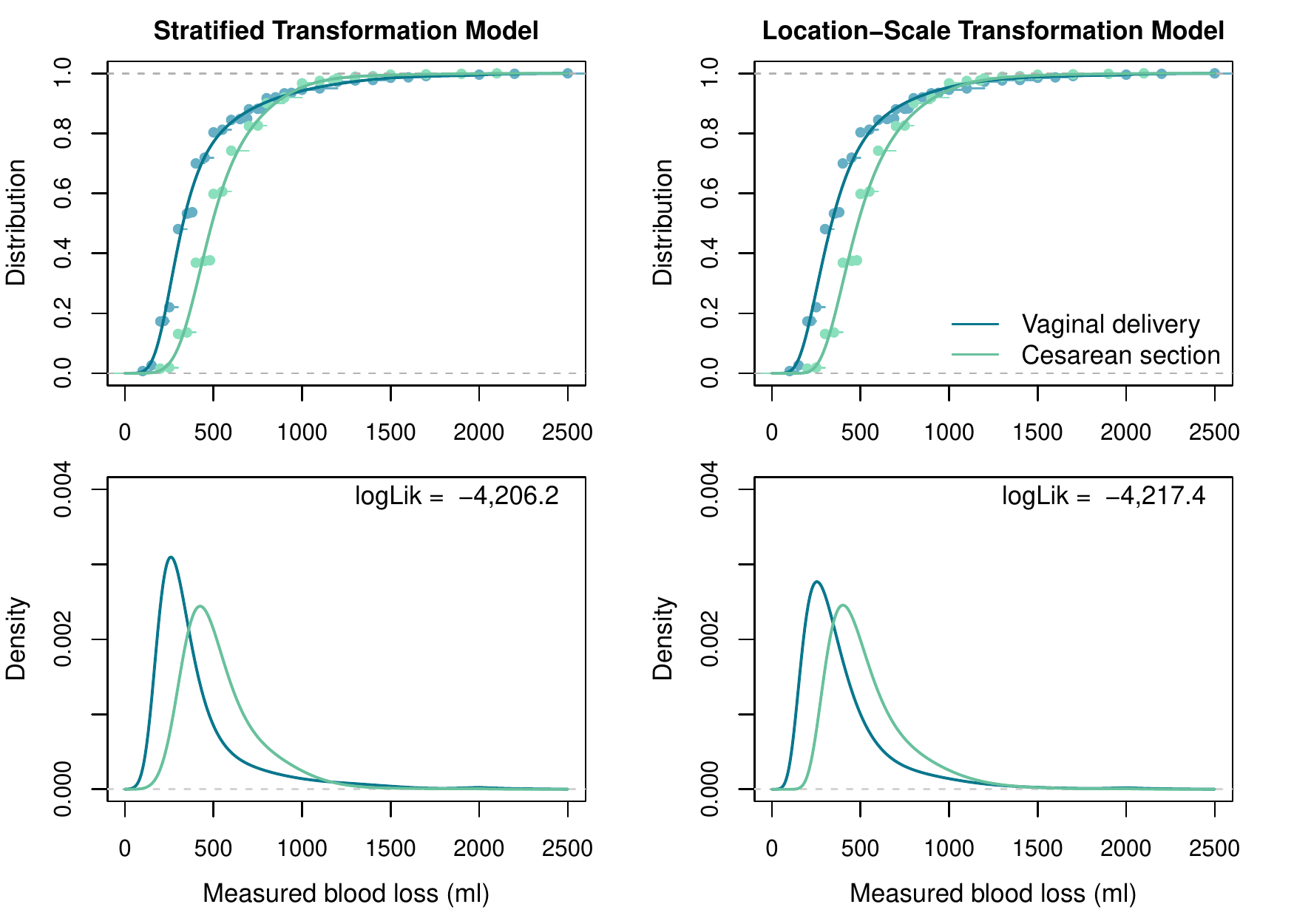} 

}

\end{knitrout}
\caption{Stratification.  Distribution (top) and density (bottom)
of postpartum blood loss conditional on delivery mode estimated by the
stratified transformation model (left) and \ls transformation model
(right).  In addition, the empirical cumulative distribution function
is shown in the top row, in-sample log-likelihoods are given in the bottom row.
\label{fig:STRAT}}
\end{figure}

Due to the practical challenges in measuring blood loss in the hectic
environment of a delivery ward, interval-censored observations were reported and the
corresponding interval-censored negative log-likelihood~(\ref{eq:dll}) is
minimized by Augmented Lagrangian Minimization \citep{Madsen_2004}
to estimate the parameters $\parm$, $\eshiftparm$, and $\escaleparm$
simultaneously. Visual inspection of distribution and density functions as
well as the in-sample log-likelihoods in
Figure~\ref{fig:STRAT} shows that the two models are practically identical.
The two estimated conditional distribution functions cross around 1,000ml,
which is only possible due to estimation of two separate transformation functions
$\h(\ry \mid \text{delivery})$ or via the inclusion of the delivery mode dependent scale term
$\sqrt{\exp(\escaleparm)}$.

We can also compute the probabilistic index here, which indicates that 
a randomly selected woman having a vaginal delivery has a probability of
$ 0.71 $ (95\%~confidence interval:
$0.68-0.74$)
for a lower blood loss compared to a randomly selected
woman undergoing a cesarean section.

\subsubsection{Crossing hazards} \label{sec:XH}
\begin{figure}[t!]
\centering
\begin{knitrout}\small
\definecolor{shadecolor}{rgb}{1, 1, 1}\color{fgcolor}

{\centering \includegraphics[width=\textwidth]{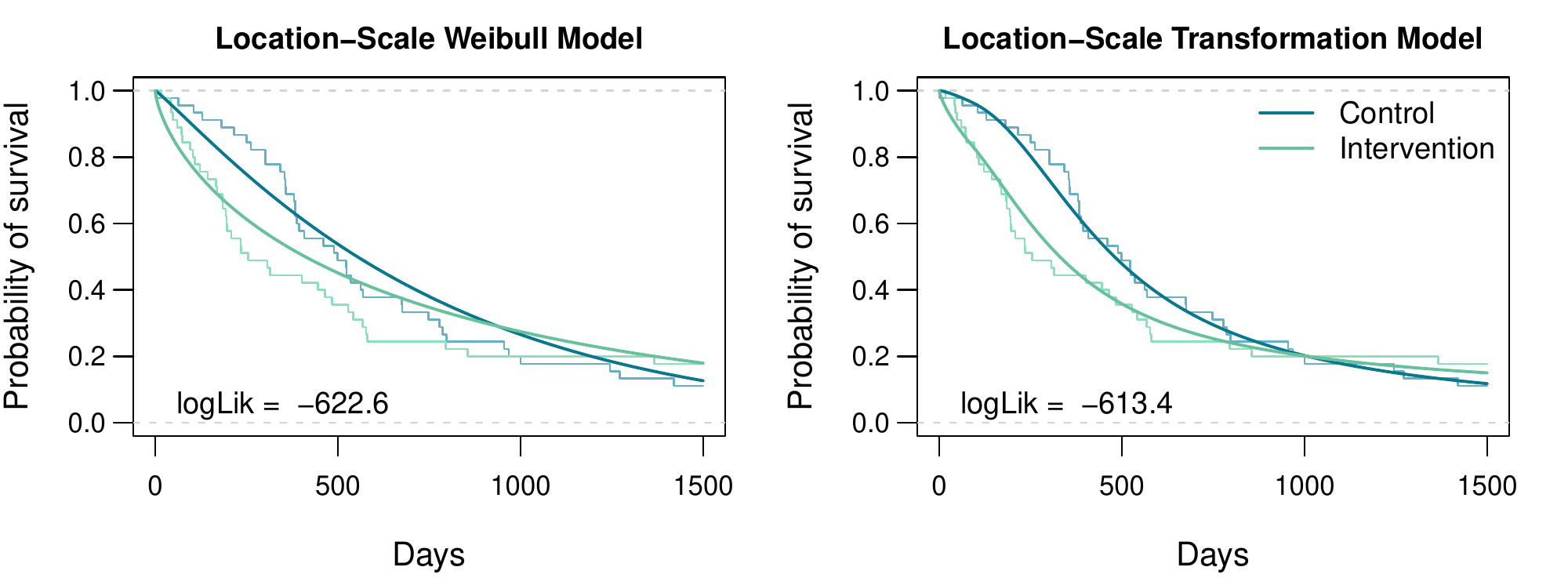} 

}

\end{knitrout}
\caption{Crossing hazards.  The survivor functions of the two groups
estimated by the non-parametric Kaplan-Meier method (step function) are shown
along the estimates from the \ls Weibull model (left) and the
distribution-free \ls transformation model (right).}
\label{fig:XH}
\end{figure}

In the following we re-analyze a two-arm randomized controlled trial of
90~patients with gastric cancer \citep{Schein_1982}. 
Trial patients received either chemotherapy
and radiotherapy (intervention group) or chemotherapy alone (control group). 
The non-parametric Kaplan-Meier estimates of the survivor functions of both
groups in Figure~\ref{fig:XH} reveal crossing of the curves at
approximately 1,000~days, and thus non-proportional hazards.

Non-proportional hazards are a common violation of a standard model assumption
in survival analysis necessitating tailored models to express differences in survival
times $T \in \RR^+$ by interpretable parameters. We suggest a \ls transformation
model of the form
\begin{eqnarray*}
\Prob(T > t \mid \text{control}) & = & \exp[-\exp\{\h(t \mid \parm)\}] \\
\Prob(T > t \mid \text{intervention}) & = &
\exp\left[-\exp\left\{\sqrt{\exp(\escaleparm)}\h(t \mid \parm) - \eshiftparm\right\}\right].
\end{eqnarray*}
For $\escaleparm = 0$, the model reduces to a proportional hazards model
with log-hazard ratio $\eshiftparm$. A distribution-free version can be
implemented by choosing a polynomial in Bernstein form for the
transformation function $\h(t \mid \parm) = \bern{6}(t)^\top
\parm$.  A Weibull model corresponds to a log-linear transformation function
$\h(t \mid \parm) = \eparm_1 + \eparm_2 \log(t)$, which was introduced
as a special case in the GAMLSS framework \citep[Table~1]{Rigby_Stasinopoulos_2005}
and later investigated in more detail by
\cite{Burke_2017} and \cite{Burke_SiM_2020} using the equivalent parameterization
$\eparm_1 + \exp(\eparm_2) \log(t)$ for the control and $\eparm_1 +
\exp(\eparm_2 + \escaleparm) \log(t) + \eshiftparm$ for the intervention
group. Further extensions of the Weibull \ls model were studied in \cite{Burke_2020}
and a non-parametric approach, leaving $\h$ completely unspecified, was
theoretically discussed by \cite{Zeng_Lin_2007} and \cite{Burke_SJoS_2020}.

Model parameters for both models were estimated by maximizing the likelihood
defined by~(\ref{eq:cll}) for death times and likelihood~(\ref{eq:dll}) for
right-censored observations.  The non-parametric Kaplan-Meier estimates in
Figure~\ref{fig:XH} are overlaid with survivor functions obtained from the
Weibull model (log-linear $\h$, left panel) and the distribution-free \ls model
($\h$ being a polynomial in Bernstein form, right panel).  Both models show
crossing survivor curves and the more flexible model appears to have better fit. 
However, in the context of a randomized trial, a test for the null of equal
survivor curves is more important than model fit. The likelihood ratio
tests lead to a rejection of the null hypothesis at $5\%$
in either model ($p$-value $= 0.034$
for the Weibull model and
$p$-value $= 0.011$ for the
distribution-free model). The bivariate Wald-test, proposed by \cite{Burke_2017} for 
crossing-hazards problems, also leads to a rejection with
$p$-value $=0.032$.

An alternative \ls test can be motivated in analogy to the
log-rank test.  The bivariate permutation score test for $\escaleparm$ and
$\eshiftparm$ for testing the null $\escaleparm = \eshiftparm = 0$ is
defined based on the unconditional model $\Prob(T > t) = \exp[-\exp\{\h(t
\mid \parm)\}]$, that is, the model fitted under the constraint $\escaleparm
= \eshiftparm = 0$.  Thus, the likelihood contribution of the $i$th subject
is $\ell_i\left(\parm, \eshiftparm, \sqrt{\exp(\escaleparm)}^{-1}\right) = \ell_i(\parm, 0, 1)$. 
The maximum-likelihood estimator is
\begin{eqnarray*}
\hat{\parm} = \argmax_{\parm} \sum^{N}_{i = 1} \ell_i(\parm, 0, 1),
  \quad \text{subject to } \eparm_p \le \eparm_{p + 1}, p \in 1, \dots, \dimparm - 1.
\end{eqnarray*}
The individual score contributions are defined as
\begin{eqnarray*}
\rr_i = \left.\frac{\partial \ell_i\left(\parm, \eshiftparm, \sqrt{\exp(\escaleparm)}^{-1}\right)}
{\partial(\eshiftparm, \escaleparm)}\right|_{\parm = \hat{\parm},
\eshiftparm = 0, \escaleparm = 0} \in \RR^2.
\end{eqnarray*}
Note that the first element of $\rr_i$ is the log-rank score for the $i$th
individual and a bivariate linear test statistic is simply the sum of the
scores in the intervention group.  After appropriate standardization,
maximum-type statistics or quadratic forms can be used to obtain $p$-values
from the asymptotic or approximate permutation distribution
\citep{Hothorn_etal_2006}.  The log-rank test alone does not lead to a
rejection at $5\%$ ($p$-value $=0.638$) but the bivariate test does
($p$-value $=0.002$ for the maximum-type and
$p$-value $=0.001$ for the quadratic form).

\subsubsection{Partial proportional hazards} \label{sec:PPH}

We analyze a time series of daily deer-vehicle collisions (DVCs) involving roe
deer that were documented over a period of ten years (2002 -- 2011) in Bavaria, Germany
\citep{Hothorn_Mueller_Held_2015}. In total, 341,655~DVCs
were reported over 3,652 days, with daily counts 
$16 \le \rY \le 210$.

As a benchmark, we fitted a location transformation model
\begin{eqnarray*}
\Prob(\rY > \ry \mid \text{day} = d, \text{year}) = \exp\left[-\exp\left\{
  \h(\lfloor \ry \rfloor \mid \parm) - \eshiftparm_\text{year} -
  \eshiftparm_{\text{weekday}(d)} - s(d \mid \shiftparm)\right\}\right]
\end{eqnarray*}
featuring log-hazard ratios for the year (baseline 2002) and day of week
(baseline Monday) and a seasonal effect $s(d \mid \shiftparm)$ 
modeled as a superposition of sinusoidal waves of different frequencies
\citep[this is a simplification of a model discussed in][]{Siegfried_Hothorn_2020}.
Two \ls models expressing $\Prob(\rY > \ry \mid \text{day} = d,
\text{year})$ by
\begin{eqnarray*}
\exp\left[-\exp\left\{
  \sqrt{\exp(s(d \mid \scaleparm))} \h(\lfloor \ry \rfloor \mid \parm) - \eshiftparm_\text{year} -
  \eshiftparm_{\text{weekday}(d)} - s(d \mid \shiftparm)\right\}\right]
\end{eqnarray*}
were additionally estimated. Because the year does not affect the scale term,
parameters $\eshiftparm_\text{year}$ are interpretable as log-hazard ratios
common to all days $1, \dots, 365$ within a year. In this sense, the model
is a partial proportional hazards model. As in
Section~\ref{sec:XH}, we study a distribution-free version ($\h$ in
Bernstein form) and a more restrictive Weibull model (log-linear $\h$)
which, for counts, is
applied to the greatest integer $\lfloor \ry \rfloor$ less than or equal to the
cut-off point $\ry$. The
correct interval-censored likelihood~(\ref{eq:dll}) for count data was used in the three cases
to estimate the unknown model parameters $\parm$, $\scaleparm$,
$\eshiftparm_\text{year}$, $\eshiftparm_{\text{weekday}(d)}$ and
$\shiftparm$ simultaneously \citep{Siegfried_Hothorn_2020}
by Augmented Lagrangian Minimization \citep{Madsen_2004}.

\begin{figure}
\centering
\begin{knitrout}\small
\definecolor{shadecolor}{rgb}{1, 1, 1}\color{fgcolor}

{\centering \includegraphics[width=.9\textwidth]{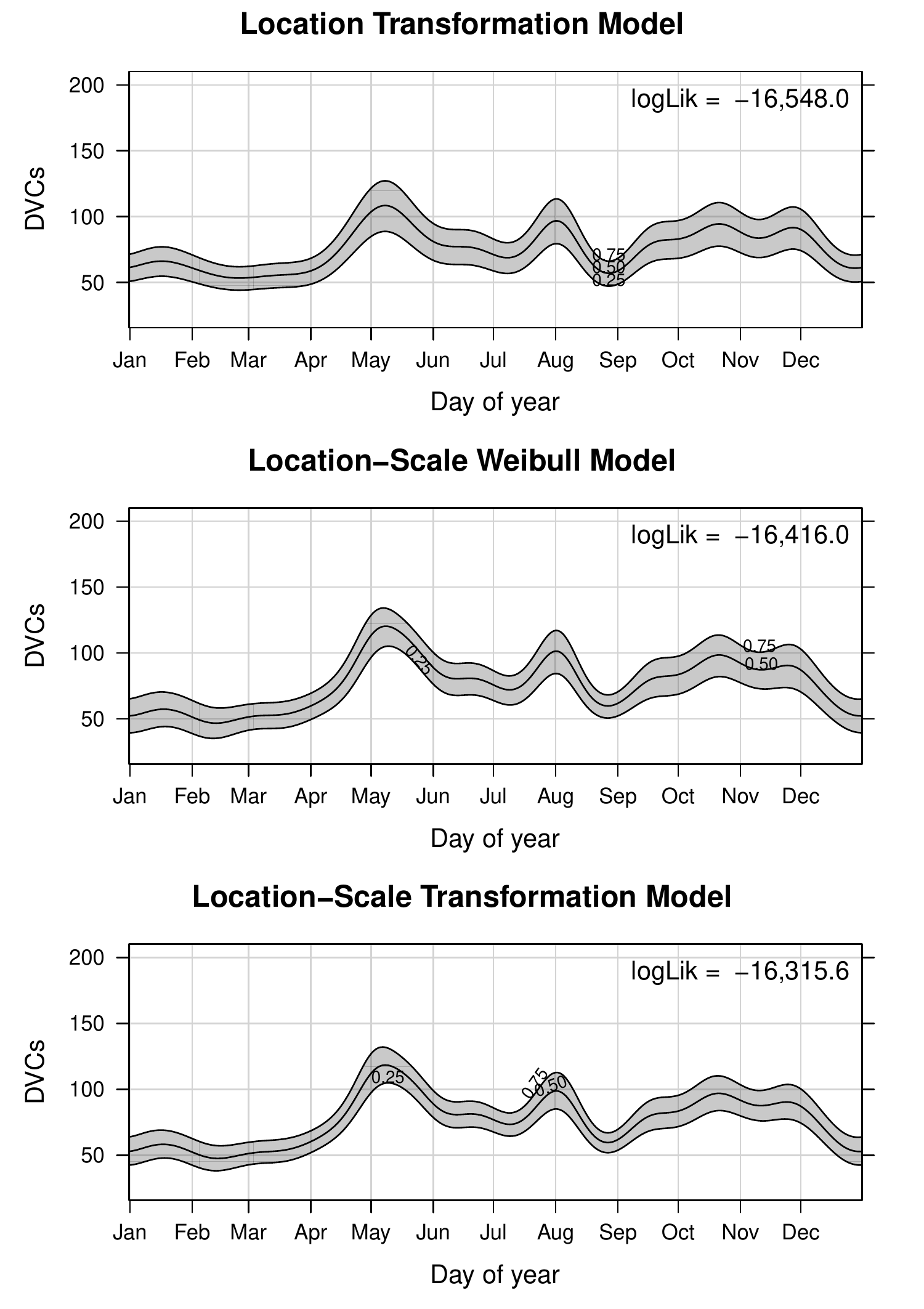} 

}

\end{knitrout}
\vspace*{-3mm}
\caption{Partial proportional hazards. Three annual quantile functions
($0.25$, $0.50$, and $0.75$th quantile)
for DVCs (for a hypothetical Monday in~2002) estimated by three
transformation models of increasing complexity. The in-sample
log-likelihoods of the corresponding models are given in the panels.}
\label{fig:PPH}
\end{figure}

Assessing the temporal changes in DVC risk across a year, all three models are
displayed on the quantile scale in Figure~\ref{fig:PPH} for a hypothetical Monday in 2002.
The curves reveal well-known seasonal patterns of increased DVC risk 
in April, July, and August due to increased animal activity.
The plots further indicate that for the location
transformation model large median values are associated with larger
dispersion.  This is not the case for the other two models, indicating a
certain degree of underdispersion. The median annual risk pattern is very similar
in all three models, however, the distribution-free \ls transformation model
reveals smaller variance compared to the other two models.

\begin{table}
\caption{Partial proportional hazards. Estimates and corresponding 
         simultaneous 95\%~confidence intervals (CI) of multiplicative changes 
         in hazards by year. Hazard ratios smaller one indicate increasing DVCs when
         comparing two subsequent years. \label{tab:PPH}}
\begin{tabular}{lrr}
  \toprule
Year & Hazard ratio & 95\% CI \\ 
  \midrule
2003 -- 2002 & 0.66 & 0.53 -- 0.81 \\ 
  2004 -- 2003 & 0.74 & 0.60 -- 0.91 \\ 
  2005 -- 2004 & 0.92 & 0.75 -- 1.13 \\ 
  2006 -- 2005 & 1.12 & 0.91 -- 1.38 \\ 
  2007 -- 2006 & 0.58 & 0.47 -- 0.72 \\ 
  2008 -- 2007 & 0.88 & 0.72 -- 1.09 \\ 
  2009 -- 2008 & 0.99 & 0.80 -- 1.21 \\ 
  2010 -- 2009 & 0.99 & 0.80 -- 1.21 \\ 
  2011 -- 2010 & 1.08 & 0.88 -- 1.32 \\ 
   \bottomrule
\end{tabular}

\end{table}

The partial proportional hazards \ls transformation model further allows
investigation of the general trend of
DVCs over a decade. From the log-hazard ratios $\eshiftparm_\text{year}$ we
computed multiple comparisons of hazard ratios comparing subsequent years,
with multiplicity control. Table~\ref{tab:PPH} is in line with an increasing
DVC risk from 2002 to 2004, followed by a plateau in 2005 and 2006, a
further risk increase in 2007, and then plateauing in the remaining years.
 

\subsection{Location-scale transformation trees}\label{sec:TRTF}

\cite{Pollet_Nettle_2009} analyzed the self-reported orgasm
frequency of 1,533~Chinese women with current male
partners.  The ordinal outcome $\rY$ was reported in terms of ordered categories:
never $<$ rarely $<$ sometimes $<$ often $<$ always.  To assess the effect
of explanatory variables on the distribution of reported orgasm frequencies, we
re-analyze this data using a tree-structured \ls transformation
model.  Explanatory variables included in the model are: partner income, partner
height, duration of the relationship, respondents age (Rage), difference between
education and wealth between both partners, the respondents education
(Reducation: no school $<$ primary school $<$ lower-middle school $<$ upper-middle school $<$ junior college $<$ university), health,
happiness (Rhappy: very unhappy $<$ not too unhappy $<$ relatively happy $<$ very happy)
and place of living (Rregion). 

\begin{figure}[t!]
\centering
\begin{knitrout}\small
\definecolor{shadecolor}{rgb}{1, 1, 1}\color{fgcolor}

{\centering \includegraphics[width=\textwidth]{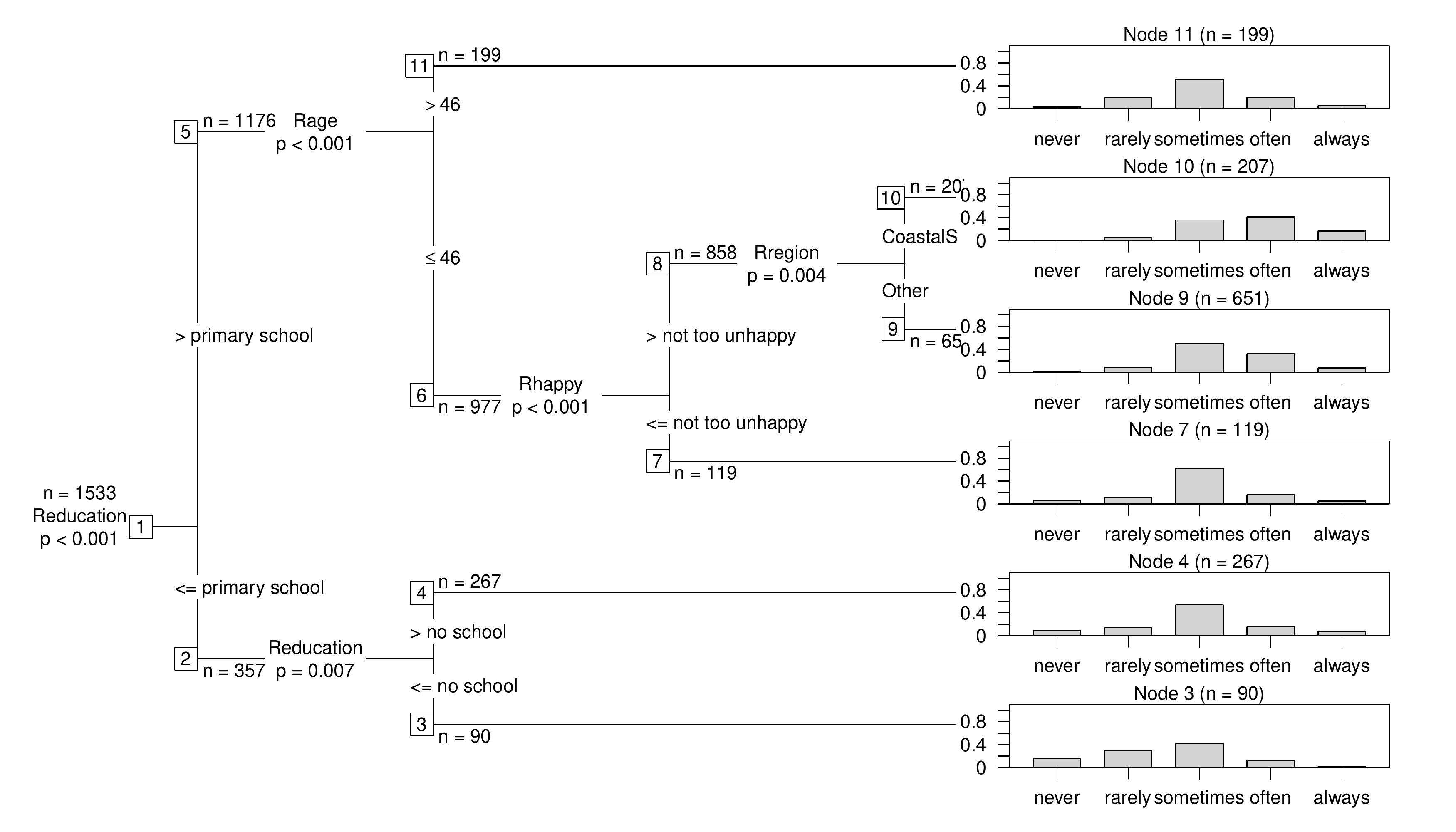} 

}

\end{knitrout}
\caption{\Ls transformation tree. Female orgasm frequency in heterosexual
relationships as a function of questionnaire variables reported by the female
respondent. }
\label{fig:TRTF}
\end{figure}

We apply a modification of the transformation tree induction algorithm
by \cite{Hothorn_Zeileis_2017} to estimate the \ls transformation tree: (i) Fit an unconditional
transformation model, (ii) assess the correlation of model scores and
explanatory variables, (iii) find an appropriate binary split in the explanatory
variable strongest correlated to the scores, (iv) proceed recursively. The
novelty here is that \ls trees pay attention to bivariate \ls scores
exclusively, instead of the $\dimparm$ scores for the transformation
parameters $\parm$ \citep[as in][]{Hothorn_Zeileis_2017}.
As in Section~\ref{sec:XH}, the unconditional model
\begin{eqnarray*}
\Prob(\rY \le \ry) = \pZ\left(\sqrt{\exp(\escaleparm)} \basisy(\ry)^\top \parm - \eshiftparm\right),
\quad \text{subject to } \eshiftparm = \escaleparm = 0
\end{eqnarray*}
is fitted in each node of the tree by optimizing the likelihood
\begin{eqnarray*}
\hat{\parm} = \argmax_{\parm} = \sum^{N}_{i = 1} \ell_i(\parm, 0, 1),
\quad \text{subject to } \eparm_p \le \eparm_{p + 1}, p \in 1, \dots, \dimparm - 1.
\end{eqnarray*}
The bivariate score contributions are defined by 
\begin{eqnarray*}
\rr_i = \left.\frac{\partial \ell_i\left(\parm, \eshiftparm,
\sqrt{\exp(\escaleparm)}^{-1}\right)}{\partial(\eshiftparm,
\escaleparm)}\right|_{\parm = \hat{\parm}, \eshiftparm = 0, \escaleparm = 0} \in \RR^2
\end{eqnarray*}
Permutation tests are then applied to assess the association between the
$j$th explanatory variable based on a quadratic form collapsing the
linear test statistic $\sum^{N}_{i=1} g_j(\rx_i) \rr_i^\top \in
\RR^{Q(j) \times 2}$, where $g_j(\rx_i)$ is a $Q(j)$-dimensional vector
representing the $j$th explanatory variable of the $i$th subject. The
bivariate score allows the tree to detect location and scale effects, for the
model in Figure~\ref{fig:TRTF} on the logit scale. A
$p$-value is computed for all $j = 1, \dots, J$ explanatory variables and
the variable with minimum $p$-value is selected for splitting.

The \ls transformation tree (Figure~\ref{fig:TRTF}) indicates that higher orgasm frequencies were in
general reported from higher educated, happier, and younger females. In this subgroup,
the coastal south region was associated with a tendency to higher reported
orgasm frequencies compared to the rest of China.

\subsection{Transformation additive models for location and scale} \label{sec:TAMLS}
\begin{figure}[t!]
\begin{knitrout}\small
\definecolor{shadecolor}{rgb}{1, 1, 1}\color{fgcolor}

{\centering \includegraphics[width=.9\textwidth]{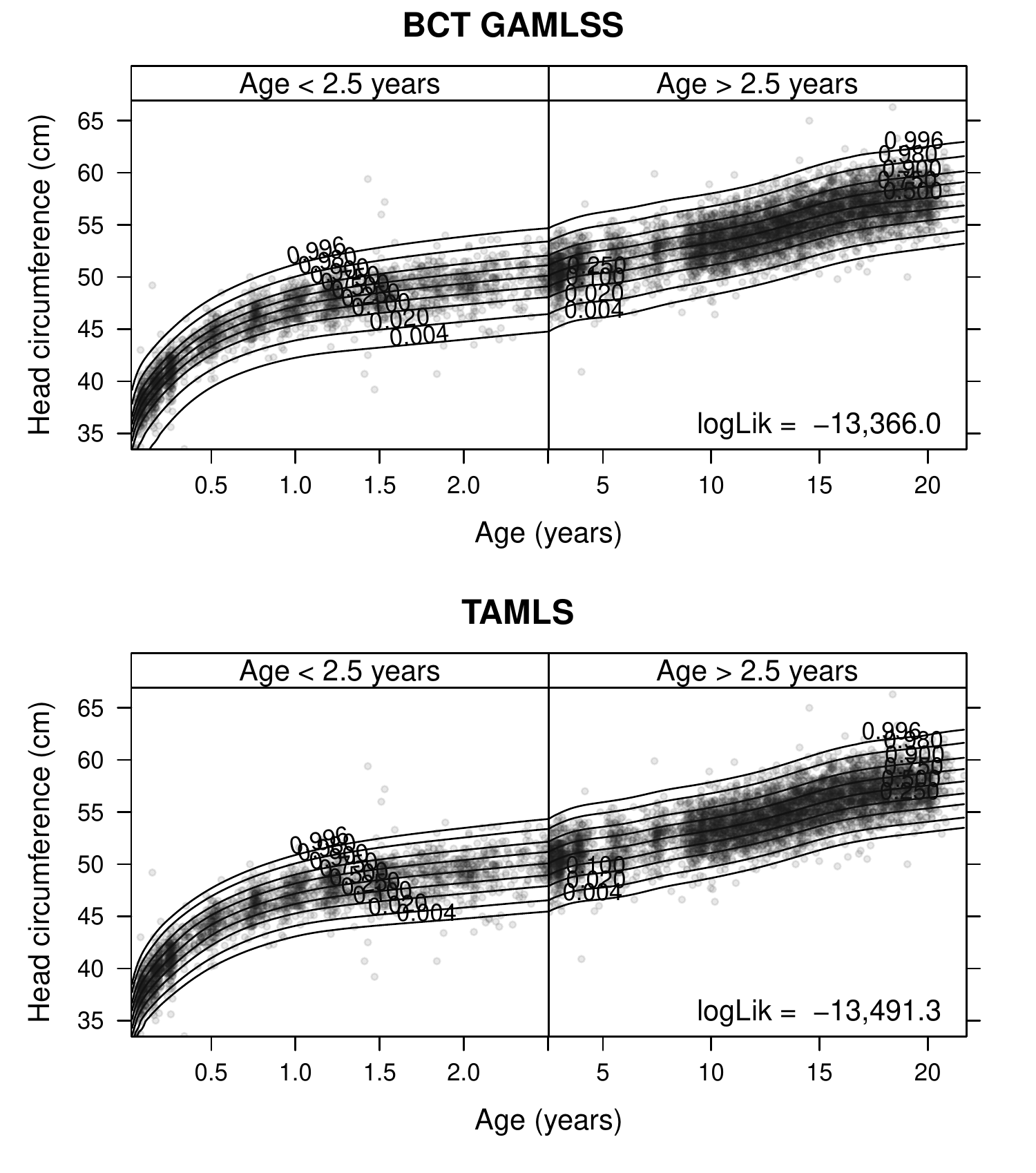} 

}

\end{knitrout}
\vspace*{-3mm}
\caption{Transformation additive models for location and scale (TAMLS). 
Conditional quantiles of head circumference along age estimated by the
Box-Cox-$t$ GAMLSS (BCT GAMLSS, top panel) and the TAMLS (bottom
panel).  The former model comprises four and the latter model two smooth
terms.\label{fig:TAMLS}}
\end{figure}

The head-circumference growth chart obtained from the Dutch growth
study \citep{Fredriks_Buuren_Burgmeijer_2000} is one of the standard
examples in the GAMLSS literature.  The top panel of Figure~\ref{fig:TAMLS}
shows the head-circumference quantiles for boys conditional on age obtained
from fitting a GAMLSS with Box-Cox-$t$ distribution, featuring four model
terms $\mu(\text{age}), \sigma(\text{age}), \nu(\text{age})$ and
$\tau(\text{age})$ \citep[reproducing Figure~16 in][]{GAMLSS_JSS_2007}.  In
our re-analysis, we replace the four parameter Box-Cox-$t$ GAMLSS with
a distribution-free transformation additive model for location and scale (TAMLS) 
featuring a conditional distribution function
\begin{eqnarray*}
\Prob(\rY \le \ry \mid \text{Age} = \text{age}) =
\Phi\left(\sigma(\text{age})^{-1}\bern{6}(\ry)^\top \parm - \mu(\text{age})\right)
\end{eqnarray*}
for head-circumference $\rY \in \RR^+$.

In contrast to the GAMLSS, there is no need to assume a specific parametric distribution in the TAMLS and only two instead
of four smooth terms have to be estimated.  In this model, the transformation parameters
$\parm$ can be understood as nuisance parameters.  We employ the Rigby and
Stasinopoulos (RS) algorithm \citep{Rigby_Stasinopoulos_2005} developed for
GAMLSS to estimate the two smooth terms $\mu(\text{age})$ and
$\sigma(\text{age})$ in our TAMLS.  For a given likelihood depending on a \l
and scale term, this algorithm allows estimation of these two terms in a
structured additive way. We shield the more complex formulation of our model
from the RS algorithm by setting-up a profile likelihood which, under the
hood, estimates the nuisance parameters
$\parm$ given $\mu$ and $\sigma$ controlled by the RS algorithm.  More
specifically, for candidate functions $\mu$ and $\sigma$, the profile likelihood
over $\parm$ is given by
\begin{eqnarray*}
\ell(\mu(\cdot), \sigma(\cdot)) = \argmax_{\parm}
   \sum_{i = 1}^N \ell_i(\parm, \mu(\rx_i), \sigma(\rx_i)) \quad \text{s.t. }
   \eparm_p \le \eparm_{p + 1}, p \in 1, \dots, \dimparm - 1.
\end{eqnarray*}
We used log-likelihood contributions~(\ref{eq:cll}) in this specific
application. Augmented Lagrangian Minimization \citep{Madsen_2004} was used
to estimate $\parm$ given $\mu(\cdot)$ and $\sigma(\cdot)$.
The penalized profile likelihood was optimized with respect to the two functions
$\mu(\cdot)$ and $\sigma(\cdot)$ in Step~2a(i) of the RS~algorithm
\citep[Appendix~B,][]{Rigby_Stasinopoulos_2005}.
The in-sample log-likelihood of the four term Box-Cox-$t$ GAMLSS is
slightly larger than the one of the distribution-free TAMLS,
but the conditional quantile sheets obtained from the two models are
very close and hardly distinguishable for boys older than $2.5$ years
(Figure~\ref{fig:TAMLS}).

Models assuming additivity of multiple smooth terms for the location effect
$\mu(\rx) = \sum_{j = 1}^J m_j(\rx)$ and the scale effect
$-2\log(\sigma(\rx)) = \sum_{l = 1}^L s_l(\rx)$ can be fitted by maximizing the
same profile likelihood using the RS~algorithm or $L_2$ boosting \citep[for
GAMLSS,][]{Mayr_Fenske_Hofner_2012}.  In this sense, transformation models
introduce a novel distribution-free member to the otherwise strictly
parametric GAMLSS family.
 

\subsection{Model selection} \label{sec:VS}

In the following we aim to assess the effect of explanatory variables on the
medical demand by the elderly, \ie number of physician visits $\rY = 0, 1, 2,
\dots $ for patients aged 66 or older, using a sample from the United States
National Medical Expenditure Survey conducted in 1987 and 1988
\citep{Deb_Trivedi_1997}.

%
For such applications, \ls transformation models~(\ref{eq:model}) are
especially attractive for parameter interpretation when linear \l
and scale terms are considered, and variables of special interest are present
in the \l term only (Section~\ref{sec:inter}).  If continuous explanatory
variables $\rx$ are present in the model, a parameter identification issue
arises which has previously been discussed in the GAMLSS context
\citep[][page 60]{GAMLSS_2019}. In a \ls model,
\begin{eqnarray*}
\Prob(\rY \le \ry \mid \rX = \rx) & = & \pZ\left(\sqrt{\exp(\rx^\top\scaleparm)}
\h(\ry \mid \parm) - \rx^\top \shiftparm\right)
\end{eqnarray*}
the intercept, which is implicit in the transformation function $\h(\ry \mid \parm) =
\bar{\h}(\ry \mid \bar{\parm}) - \eshiftparm_0$, must not be multiplied with
the scale term and an explicit intercept must be added to the \l term,
changing the model to
\begin{eqnarray} \label{eq:modelintercept}
\Prob(\rY \le \ry \mid \rX = \rx) & = & \pZ\left(\sqrt{\exp(\rx^\top\scaleparm)} \bar{\h}(\ry \mid
\bar{\parm}) - \eshiftparm_0 - \rx^\top
\shiftparm\right).
\end{eqnarray}
The two models are not equivalent, but adding $\eshiftparm_0$ to $\h$ leads to
an unidentified parameter when $\scaleparm$ is close to zero and omitting
$\eshiftparm_0$ leads to different model fits when a constant is added to a
continuous explanatory variable (\eg~when defining a suitable
baseline distribution).  For discrete parameterizations the expression for
$\bar{\h}$ simplifies to $\bar{\h}(\ry \mid \parm) = \basisy(\ry)^\top
\parm$ with $\eparm_1 \equiv 0$.  For polynomials in Bernstein form we have
the following expression for $\h(\ry \mid \parm)$:
\begin{eqnarray*}
\bern{\dimparm - 1}(\ry)^\top \parm =
\sum_{p = 1}^P a_p(\ry) \eparm_p = \sum_{p = 1}^{P - 1} (a_p(\ry) -
a_P(\ry)) \bar{\eparm}_p + \dimparm^{-1} \sum_{p = 1}^P \eparm_p
= \bar{\h}(\ry \mid \bar{\parm}) - \eshiftparm_0
\end{eqnarray*}
and $\int_\RR \bar{\h}(\ry \mid \bar{\parm}) \, d\ry = 0$ because
$a_p(\cdot)$ are densities.  The model parameters $\bar{\parm}$,
$\scaleparm$, and $\shiftparm$ can be estimated by maximizing the likelihood
(after suitable adjustment to the constraints).  However, for the sake of
interpretability we aim to drop variables from the scale term whenever
possible and therefore apply the $L_0$ penalty (detailed in Section~\ref{sec:modsel},
implemented in package~\pkg{tramvs}~\citep{pkg:tramvs})
on $\scaleparm$ to the likelihood of model~(\ref{eq:modelintercept}).
%

\begin{table}
\caption{Model selection. \L and scale parameter estimates, $\hat\eshiftparm$ and $\hat\escaleparm$,
from applying the two estimation procedures, maximum likelihood (ML) or best subset selection (BSS),
to a \ls transformation model including the following explanatory variables:
Health (poor $<$ average (baseline) $<$ excellent),
Sex (female (baseline), male), Insurance (no (baseline), yes), 
Chronic (number of chronic conditions) and School (number of years of education).
Variables which were dropped when applying best subset selection are indicated by ---.
\label{tab:VS}}
\begin{tabular}{llrrrr}
  \toprule &&\multicolumn{2}{c}{ML}&\multicolumn{2}{c}{BSS}\\\cmidrule(l){3-4} \cmidrule(l){5-6}Variable & Level & $\hat{\beta}$ & $\hat{\gamma}$ & $\hat{\beta}$ & $\hat{\gamma}$ \\ 
  \midrule
Health & poor & $ 0.3315 $ & $ -0.0912 $ & $ 0.3997 $ & --- \\ 
   & excellent & $ -0.3722 $ & $ 0.0417 $ & $ -0.2214 $ & --- \\ 
  Sex & male & $ -0.1585 $ & $ -0.2669 $ & $ -0.1251 $ & --- \\ 
  Insurance & yes & $ 0.2675 $ & $ 0.2541 $ & $ 0.2800 $ & $ 0.2272 $ \\ 
  Chronic &  & $ 0.2542 $ & $ 0.1799 $ & $ 0.2729 $ & $ 0.2260 $ \\ 
  School &  & $ 0.0234 $ & $ -0.0175 $ & $ 0.0233 $ & --- \\ 
   \bottomrule
\end{tabular}

\end{table}

Applying the two estimation procedures, maximum likelihood and best subset
selection, to a \ls transformation model (with $F = \cloglog^{-1}$)
estimating the effect of self-perceived health status (Health), sex (Sex),
insurance coverage (Insurance), and the number of chronic conditions
(Chronic) and years of education of patients (School) on the frequency of
physician visits, allows for a head-to-head comparison of the parameter
estimates (Table~\ref{tab:VS}).  In the best subset \ls transformation model
the variables Health, Sex and School are dropped from the scale term
allowing to interpret their effects in terms of (log-)hazard ratios.  For
the variable Sex, for example, the corresponding $\exp(-\hat\eshiftparm) =
1.1333$ can be interpreted as
hazard ratio comparing the hazards of male patients to the hazards of female
patients, all other variables being equal.

\section{Discussion}

\cite{Tosteson_1988} introduced the notion of distribution-free \ls
regression in the context of ROC analysis.  While they were able to estimate
a corresponding model for ordinal responses, they contemplated that for
models~(\ref{eq:model}), ``there is, as yet, no software for accommodating
continuous test results, which are common outcomes for laboratory tests''
\citep{Tosteson_1988}.  With the introduction of a smooth transformation
function and corresponding software implementation in the \pkg{tram} add-on
package \citep{pkg:tram}, we address this long-standing issue.  
%
%
We derive likelihood and score contributions for all response types
and discuss suitable inference procedures for various functional forms.

In a broader context, we contribute a new distribution-free member to 
the rich family of \ls models. The model is unique in the sense that data
analysts do not have to commit themselves to a parametric family of
distributions before fitting the model. The flexibility of our approach
comes from the pair of \l and scale terms allowing interpretability of
conditional distributions on various scales, including proportional odds or
hazards. Despite the distribution-free nature, we parameterize the model
such that simple maximum-likelihood estimation for all types of responses
becomes feasible.  Therefore, our  implementation handles arbitrary responses, including
bounded, mixed discrete-continuous, and randomly censored outcomes, in a
native way. Among other
diverse applications, our flexible approach can help to generalize Weibull \ls
models previously studied as a model for crossing hazards using GAMLSS,
allows for over- or underdispersion to be explained by covariates in complex
count regression models, adds a notion of dispersion to regression trees for
complex responses, provides means to reduce the complexity of growth-curve
models, and has important applications in ROC analysis \citep{Sewak_Hothorn_2022}.

Special care with respect to parameter interpretability is needed when
formulating the model.  Parameters in linear location terms are
interpretable as log-odds or log-hazard ratios for as long as there is no
corresponding scale parameter.  Thus, model selection becomes vitally
important should the data analyst be interested in direct parameter
interpretation.  A novel approach to best subset selection was presented and
empirically evaluated.  Model interpretation is possible on other scales
(\eg probabilistic indices or conditional quantiles), yet constitutes
probably the biggest challenge of \ls transformation models.

All models discussed here are ``distributional'' in the sense that they
formulate a proper distribution function.  Via appropriate constraints, our
software implementation ensures that fitted models also directly correspond
to conditional distribution functions.  This feature allows straightforward
parametric bootstrap implementations.  Alternative suggestions for \ls
ordinal regression do not necessarily lead to estimates which can be
interpreted on the probability scale
\citep{CCox_1995,Tutz_Berger_2017,Tutz_Berger_2020}.

Algorithmically, we stand on the shoulders of giants, because only minor
modifications to well-established algorithms were necessary for enabling
parameter estimation.  We didn't fully explore all possibilities here, and
for example \ls transformation forests building on \cite{Hothorn_Zeileis_2017} or
functional gradient boosting for this class \citep{Hothorn_2019} are
interesting algorithms for smooth interaction modelling in potentially
high-dimensional covariate spaces.
 
\section*{Author contributions}

SS and TH developed the model and its parameterization. 
SS wrote the manuscript, analysed all applications, and performed the
simulation study. LK implemented best subset regression for linear \ls
transformation models. TH implemented the likelihood and score function in
package \pkg{mlt}. All authors revised and approved the final version of the manuscript.

\section*{Acknowledgement}

SS and TH acknowledge financial support by Swiss National Science
Foundation, Grant No.~200021\_184603. The authors would like to thank
Christoph Blapp and Klaus Steigmiller for embedding \ls transformation
models into the literature and for a proof-of-concept implementation.

\bibliography{mlt,ss,packages}

\begin{thebibliography}{42}
\newcommand{\enquote}[1]{``#1''}
\providecommand{\natexlab}[1]{#1}
\providecommand{\url}[1]{\texttt{#1}}
\providecommand{\urlprefix}{URL }
\expandafter\ifx\csname urlstyle\endcsname\relax
  \providecommand{\doi}[1]{doi:\discretionary{}{}{}#1}\else
  \providecommand{\doi}{doi:\discretionary{}{}{}\begingroup
  \urlstyle{rm}\Url}\fi
\providecommand{\eprint}[2][]{\url{#2}}

\bibitem[{Burke \emph{et~al.}(2020{\natexlab{a}})Burke, Eriksson, and
  Pipper}]{Burke_SJoS_2020}
Burke K, Eriksson F, Pipper CB (2020{\natexlab{a}}).
\newblock \enquote{Semiparametric Multiparameter Regression Survival Modeling.}
\newblock \emph{Scandinavian Journal of Statistics}, \textbf{47}(2), 555--571.
\newblock \doi{10.1111/sjos.12416}.

\bibitem[{Burke \emph{et~al.}(2020{\natexlab{b}})Burke, Jones, and
  Noufaily}]{Burke_2020}
Burke K, Jones MC, Noufaily A (2020{\natexlab{b}}).
\newblock \enquote{A Flexible Parametric Modelling Framework for Survival
  Analysis.}
\newblock \emph{Journal of the Royal Statistical Society: Series~C (Applied
  Statistics)}, \textbf{69}(2), 429--457.
\newblock \doi{10.1111/rssc.12398}.

\bibitem[{Burke and MacKenzie(2017)}]{Burke_2017}
Burke K, MacKenzie G (2017).
\newblock \enquote{Multi-Parameter Regression Survival Modeling: An Alternative
  to Proportional Hazards.}
\newblock \emph{Biometrics}, \textbf{73}(2), 678--686.
\newblock \doi{10.1111/biom.12625}.

\bibitem[{Cox(1995)}]{CCox_1995}
Cox C (1995).
\newblock \enquote{Location-Scale Cumulative Odds Models for Ordinal Data: A
  General Non-Linear Model Approach.}
\newblock \emph{Statistics in Medicine}, \textbf{14}(11), 1191--1203.
\newblock \doi{10.1002/sim.4780141105}.

\bibitem[{Deb and Trivedi(1997)}]{Deb_Trivedi_1997}
Deb P, Trivedi PK (1997).
\newblock \enquote{Demand for Medical Care by the Elderly: A Finite Mixture
  Approach.}
\newblock \emph{Journal of Applied Econometrics}, \textbf{12}, 313--336.
\newblock \doi{10.1002/(sici)1099-1255(199705)12:3<313::aid-jae440>3.0.co;2-g}.

\bibitem[{Farouki(2012)}]{Farouki_2012}
Farouki RT (2012).
\newblock \enquote{The {Bernstein} Polynomial Basis: A Centennial
  Retrospective.}
\newblock \emph{Computer Aided Geometric Design}, \textbf{29}(6), 379--419.
\newblock \doi{10.1016/j.cagd.2012.03.001}.

\bibitem[{Fredriks \emph{et~al.}(2000)Fredriks, van Buuren, Burgmeijer,
  Meulmeester, Beuker, Brugman, Roede, Verloove-Vanhorick, and
  Wit}]{Fredriks_Buuren_Burgmeijer_2000}
Fredriks AM, van Buuren S, Burgmeijer RJF, Meulmeester JF, Beuker RJ, Brugman
  E, Roede MJ, Verloove-Vanhorick SP, Wit J (2000).
\newblock \enquote{Continuing Positive Secular Growth Change in {The}
  {Netherlands} 1955--1997.}
\newblock \emph{Pediatric Research}, \textbf{47}(3), 316--323.
\newblock \doi{10.1203/00006450-200003000-00006}.

\bibitem[{Haslinger \emph{et~al.}(2020)Haslinger, Korte, Hothorn, Brun,
  Greenberg, and Zimmermann}]{Haslinger_Korte_Hothorn_2020}
Haslinger C, Korte W, Hothorn T, Brun R, Greenberg C, Zimmermann R (2020).
\newblock \enquote{The Impact of Prepartum Factor {XIII} Activity on Postpartum
  Blood Loss.}
\newblock \emph{Journal of Thrombosis and Haemostasis}, \textbf{18}(6),
  1310--1319.
\newblock \doi{10.1111/jth.14795}.

\bibitem[{Hothorn(2020{\natexlab{a}})}]{Hothorn_2018}
Hothorn T (2020{\natexlab{a}}).
\newblock \enquote{Most Likely Transformations: The \textbf{mlt} Package.}
\newblock \emph{Journal of Statistical Software}, \textbf{92}(1), 1--68.
\newblock \doi{10.18637/jss.v092.i01}.

\bibitem[{Hothorn(2020{\natexlab{b}})}]{Hothorn_2019}
Hothorn T (2020{\natexlab{b}}).
\newblock \enquote{Transformation Boosting Machines.}
\newblock \emph{Statistics and Computing}, \textbf{30}, 141--152.
\newblock \doi{10.1007/s11222-019-09870-4}.

\bibitem[{Hothorn(2023{\natexlab{a}})}]{pkg:mlt}
Hothorn T (2023{\natexlab{a}}).
\newblock \emph{\pkg{mlt}: Most Likely Transformations}.
\newblock \proglang{R}~package version~1.4-7,
  \urlprefix\url{https://CRAN.R-project.org/package=mlt}.

\bibitem[{Hothorn(2023{\natexlab{b}})}]{pkg:trtf}
Hothorn T (2023{\natexlab{b}}).
\newblock \emph{\pkg{trtf}: Transformation Trees and Forests}.
\newblock \proglang{R}~package version~0.4-2,
  \urlprefix\url{https://CRAN.R-project.org/package=trtf}.

\bibitem[{Hothorn \emph{et~al.}(2023)Hothorn, Barbanti, and
  Siegfried}]{pkg:tram}
Hothorn T, Barbanti L, Siegfried S (2023).
\newblock \emph{\pkg{tram}: Transformation Models}.
\newblock \proglang{R}~package version~0.8-3,
  \urlprefix\url{https://CRAN.R-project.org/package=tram}.

\bibitem[{Hothorn \emph{et~al.}(2006)Hothorn, Hornik, van~de Wiel, and
  Zeileis}]{Hothorn_etal_2006}
Hothorn T, Hornik K, van~de Wiel MA, Zeileis A (2006).
\newblock \enquote{A {L}ego System for Conditional Inference.}
\newblock \emph{The American Statistician}, \textbf{60}(3), 257--263.
\newblock \doi{10.1198/000313006x118430}.

\bibitem[{Hothorn \emph{et~al.}(2018)Hothorn, M\"ost, and
  B\"uhlmann}]{Hothorn_Moest_Buehlmann_2017}
Hothorn T, M\"ost L, B\"uhlmann P (2018).
\newblock \enquote{Most Likely Transformations.}
\newblock \emph{Scandinavian Journal of Statistics}, \textbf{45}(1), 110--134.
\newblock \doi{10.1111/sjos.12291}.

\bibitem[{Hothorn \emph{et~al.}(2015)Hothorn, M\"uller, Held, M\"ost, and
  Mysterud}]{Hothorn_Mueller_Held_2015}
Hothorn T, M\"uller J, Held L, M\"ost L, Mysterud A (2015).
\newblock \enquote{Temporal Patterns of Deer-Vehicle Collisions Consistent with
  Deer Activity Pattern and Density Increase but not General Accident Risk.}
\newblock \emph{Accident Analysis \& Prevention}, \textbf{81}, 143--152.
\newblock \doi{10.1016/j.aap.2015.04.037}.

\bibitem[{Hothorn and Zeileis(2021)}]{Hothorn_Zeileis_2017}
Hothorn T, Zeileis A (2021).
\newblock \enquote{Predictive Distribution Modelling Using Transformation
  Forests.}
\newblock \emph{Journal of Computational and Graphical Statistics},
  \textbf{30}(4), 144--148.
\newblock \doi{10.1080/10618600.2021.1872581}.

\bibitem[{Kneib \emph{et~al.}(2023)Kneib, Silbersdorff, and
  S{\"a}fken}]{Kneib_2021}
Kneib T, Silbersdorff A, S{\"a}fken B (2023).
\newblock \enquote{Rage Against the Mean -- {A} Review of Distributional
  Regression Approaches.}
\newblock \emph{Econometrics and Statistics}, \textbf{26}, 99--123.
\newblock \doi{10.1016/j.ecosta.2021.07.006}.

\bibitem[{Kook(2023)}]{pkg:tramvs}
Kook L (2023).
\newblock \emph{\pkg{tramvs}: Optimal Subset Selection for Transformation
  Models}.
\newblock \proglang{R}~package version~0.0-4,
  \urlprefix\url{https://CRAN.R-project.org/package=tramvs}.

\bibitem[{Lepage(1971)}]{Lepage_1971}
Lepage Y (1971).
\newblock \enquote{A Combination of {Wilcoxon's} and {Ansari-Bradley's}
  Statistics.}
\newblock \emph{Biometrika}, \textbf{58}(1), 213--217.
\newblock \doi{10.2307/2334333}.

\bibitem[{Madsen \emph{et~al.}(2004)Madsen, Nielsen, and
  Tingleff}]{Madsen_2004}
Madsen K, Nielsen HB, Tingleff O (2004).
\newblock \emph{Optimization with Constraints}.
\newblock 2nd edition. Technical University of Denmark.
\newblock \urlprefix\url{http://www2.imm.dtu.dk/pubdb/p.php?4213}.

\bibitem[{Mayr \emph{et~al.}(2012)Mayr, Fenske, Hofner, Kneib, and
  Schmid}]{Mayr_Fenske_Hofner_2012}
Mayr A, Fenske N, Hofner B, Kneib T, Schmid M (2012).
\newblock \enquote{{Generalized Additive Models for Location, Scale and Shape}
  for High-dimensional Data -- {A} Flexible Approach Based on Boosting.}
\newblock \emph{Journal of the Royal Statistical Society: Series~C (Applied
  Statistics)}, \textbf{61}(3), 403--427.
\newblock \doi{10.1111/j.1467-9876.2011.01033.x}.

\bibitem[{McCullagh(1980)}]{McCullagh_1980}
McCullagh P (1980).
\newblock \enquote{Regression Models for Ordinal Data.}
\newblock \emph{Journal of the Royal Statistical Society: Series~B
  (Methodological)}, \textbf{42}(2), 109--127.
\newblock \doi{10.1111/j.2517-6161.1980.tb01109.x}.

\bibitem[{McLain and Ghosh(2013)}]{McLain_Ghosh_2013}
McLain AC, Ghosh SK (2013).
\newblock \enquote{Efficient Sieve Maximum Likelihood Estimation of
  Time-Transformation Models.}
\newblock \emph{Journal of Statistical Theory and Practice}, \textbf{7}(2),
  285--303.
\newblock \doi{10.1080/15598608.2013.772835}.

\bibitem[{Peng \emph{et~al.}(2020)Peng, MacKenzie, and Burke}]{Burke_SiM_2020}
Peng D, MacKenzie G, Burke K (2020).
\newblock \enquote{A Multiparameter Regression Model for Interval-Censored
  Survival Data.}
\newblock \emph{Statistics in Medicine}, \textbf{39}(14), 1903--1918.
\newblock \doi{10.1002/sim.8508}.

\bibitem[{Peterson and Harrell(1990)}]{Peterson_Harrell_1990}
Peterson B, Harrell FE (1990).
\newblock \enquote{Partial Proportional Odds Models for Ordinal Response
  Variables.}
\newblock \emph{Journal of the Royal Statistical Society: Series~C (Applied
  Statistics)}, \textbf{39}(2), 205--217.
\newblock \doi{10.2307/2347760}.

\bibitem[{Pollet and Nettle(2009)}]{Pollet_Nettle_2009}
Pollet TV, Nettle D (2009).
\newblock \enquote{Partner Wealth Predicts Self-Reported Orgasm Frequency in a
  Sample of {Chinese} Women.}
\newblock \emph{Evolution and Human Behavior}, \textbf{30}(2), 146--151.
\newblock \doi{10.1016/j.evolhumbehav.2008.11.002}.

\bibitem[{{\proglang{R} Core Team}(2023)}]{R}
{\proglang{R} Core Team} (2023).
\newblock \emph{\proglang{R}: A Language and Environment for Statistical
  Computing}.
\newblock \proglang{R} Foundation for Statistical Computing, Vienna, Austria.
\newblock \urlprefix\url{https://www.R-project.org/}.

\bibitem[{Rigby \emph{et~al.}(2019)Rigby, Stasinopoulos, Heller, and {De
  Bastiani}}]{GAMLSS_2019}
Rigby R, Stasinopoulos DM, Heller G, {De Bastiani} F (2019).
\newblock \emph{Distributions for Modeling Location, Scale, and Shape: Using
  {GAMLSS} in~\proglang{R}}.
\newblock Chapman \& Hall/CRC Press, Boca Raton, FL, U.S.A.
\newblock \doi{10.1201/9780429298547}.

\bibitem[{Rigby and Stasinopoulos(2005)}]{Rigby_Stasinopoulos_2005}
Rigby RA, Stasinopoulos DM (2005).
\newblock \enquote{Generalized Additive Models for Location, Scale and Shape.}
\newblock \emph{Journal of the Royal Statistical Society: Series~C (Applied
  Statistics)}, \textbf{54}(3), 507--554.
\newblock \doi{10.1111/j.1467-9876.2005.00510.x}.

\bibitem[{Schein and {Gastrointestinal Tumor Study Group}(1982)}]{Schein_1982}
Schein PS, {Gastrointestinal Tumor Study Group} (1982).
\newblock \enquote{A Comparison of Combination Chemotherapy and Combined
  Modality Therapy for Locally Advanced Gastric Carcinoma.}
\newblock \emph{Cancer}, \textbf{49}(9), 1771--1777.
\newblock
  \doi{10.1002/1097-0142(19820501)49:9<1771::aid-cncr2820490907>3.0.co;2-m}.

\bibitem[{Sewak and Hothorn(2023)}]{Sewak_Hothorn_2022}
Sewak A, Hothorn T (2023).
\newblock \enquote{Estimating Transformations for Evaluating Diagnostic Tests
  with Covariate Adjustment.}
\newblock \emph{Statistical Methods in Medical Research}.
\newblock \doi{10.1177/09622802231176030}.
\newblock Accepted for publication 2023-04-25.

\bibitem[{Siegfried \emph{et~al.}(2023)Siegfried, Barbanti, and
  Hothorn}]{pkg:cotram}
Siegfried S, Barbanti L, Hothorn T (2023).
\newblock \emph{\pkg{cotram}: Count Transformation Models}.
\newblock \proglang{R}~package version~0.4-4,
  \urlprefix\url{https://CRAN.R-project.org/package=cotram}.

\bibitem[{Siegfried and Hothorn(2020)}]{Siegfried_Hothorn_2020}
Siegfried S, Hothorn T (2020).
\newblock \enquote{Count Transformation Models.}
\newblock \emph{Methods in Ecology and Evolution}, \textbf{11}(7), 818--827.
\newblock \doi{10.1111/2041-210x.13383}.

\bibitem[{Stasinopoulos and Rigby(2007)}]{GAMLSS_JSS_2007}
Stasinopoulos DM, Rigby RA (2007).
\newblock \enquote{Generalized Additive Models for Location Scale and Shape
  ({GAMLSS}) in~\proglang{R}.}
\newblock \emph{Journal of Statistical Software}, \textbf{23}(7), 1--46.
\newblock \doi{10.18637/jss.v023.i07}.

\bibitem[{Stasinopoulos and Rigby(2023)}]{pkg:gamlss}
Stasinopoulos M, Rigby B (2023).
\newblock \emph{\pkg{gamlss}: Generalised Additive Models for Location Scale
  and Shape}.
\newblock \proglang{R}~package version~5.4-12,
  \urlprefix\url{https://CRAN.R-project.org/package=gamlss}.

\bibitem[{Thas \emph{et~al.}(2012)Thas, {De Neve}, Clement, and
  Ottoy}]{Thas_Neve_Clement_2012}
Thas O, {De Neve} J, Clement L, Ottoy JP (2012).
\newblock \enquote{Probabilistic Index Models.}
\newblock \emph{Journal of the Royal Statistical Society: Series~B (Statistical
  Methodology)}, \textbf{74}(4), 623--671.
\newblock \doi{10.1111/j.1467-9868.2011.01020.x}.

\bibitem[{Tosteson and Begg(1988)}]{Tosteson_1988}
Tosteson ANA, Begg CB (1988).
\newblock \enquote{A General Regression Methodology for {ROC} Curve
  Estimation.}
\newblock \emph{Medical Decision Making}, \textbf{8}(3), 204--215.
\newblock \doi{10.1177/0272989x8800800309}.

\bibitem[{Tutz and Berger(2017)}]{Tutz_Berger_2017}
Tutz G, Berger M (2017).
\newblock \enquote{Separating Location and Dispersion in Ordinal Regression
  Models.}
\newblock \emph{Econometrics and Statistics}, \textbf{2}, 131--148.
\newblock \doi{10.1016/j.ecosta.2016.10.002}.

\bibitem[{Tutz and Berger(2020)}]{Tutz_Berger_2020}
Tutz G, Berger M (2020).
\newblock \enquote{Non Proportional Odds Models Are Widely Dispensable --
  {S}parser Modeling Based on Parametric and Additive Location-Shift
  Approaches.}
\newblock \emph{arXiv 2006.03914}, arXiv.org E-Print Archive.
\newblock \doi{10.48550/arXiv.2006.03914}.

\bibitem[{Zeng and Lin(2007)}]{Zeng_Lin_2007}
Zeng D, Lin DY (2007).
\newblock \enquote{Maximum Likelihood Estimation in Semiparametric Regression
  Models with Censored Data.}
\newblock \emph{Journal of the Royal Statistical Society: Series~B (Statistical
  Methodology)}, \textbf{69}(4), 507--564.
\newblock \doi{10.1111/j.1369-7412.2007.00606.x}.

\bibitem[{Zhu \emph{et~al.}(2020)Zhu, Wen, Zhu, Zhang, and Wang}]{Zhu_2020}
Zhu J, Wen C, Zhu J, Zhang H, Wang X (2020).
\newblock \enquote{A Polynomial Algorithm for Best-Subset Selection Problem.}
\newblock \emph{Proceedings of the National Academy of Sciences},
  \textbf{117}(52), 33117--33123.
\newblock \doi{10.1073/pnas.2014241117}.

\end{thebibliography}

\newpage
\begin{center}
\textbf{\LARGE{Supplementary Material}}
\end{center}
\renewcommand{\thesection}{S\arabic{section}}
\renewcommand{\thefigure}{S\arabic{figure}}
\begin{appendix}

\section{Computational details}

All computations were performed using
\proglang{R}~version 4.3.0  \citep{R}.
A reference implementation of linear \ls transformation models is available
in the \proglang{R}~add-on package \pkg{tram} \citep{pkg:tram}, which
was build on the infrastructure of the \pkg{mlt} package \citep{Hothorn_2018,pkg:mlt}.
Computations were performed using \pkg{mlt}~version 1.4.7 and
\pkg{tram}~version 0.8.3 and add-on packages \pkg{cotram} \citep[version~0.4.4,][]{pkg:cotram}
for the estimation of count \ls transformation models, \pkg{trtf} \citep[version~0.4.2,][]{pkg:trtf}
for estimating tree-structured models and package \pkg{tramvs} \citep[version~0.0.4,][]{pkg:tramvs} for best subset selection.
GAMLSS were estimated using the \pkg{gamlss} \citep[version~5.4.12,][]{pkg:gamlss} package.

\section[Re-analysis of TG]{Re-analysis of \cite{Tosteson_1988}} \label{sec:ROC}

In their analysis \cite{Tosteson_1988} estimated model~(\ref{eq:TB}) to
assess the efficacy of ultrasound for the detection of hepatic metastases in
patients with primary cancers of either the breast or the colon.  The
ultrasound rating was reported in ordered categories $\rY \in \{1 < 2 < 3 <
4 < 5\}$.  Model~(\ref{eq:TB}) with $\pZ = \Phi$ was fitted to the
reported categories, including disease status (presence or absence of hepatic
metastases) and tumor site (breast or colon), and an interaction thereof, in
the location term $\mu(\rx)$ and scale $\sigma(\rx)$ term.

\begin{figure}[t!]
\centering
\includegraphics[width = .45\textwidth]{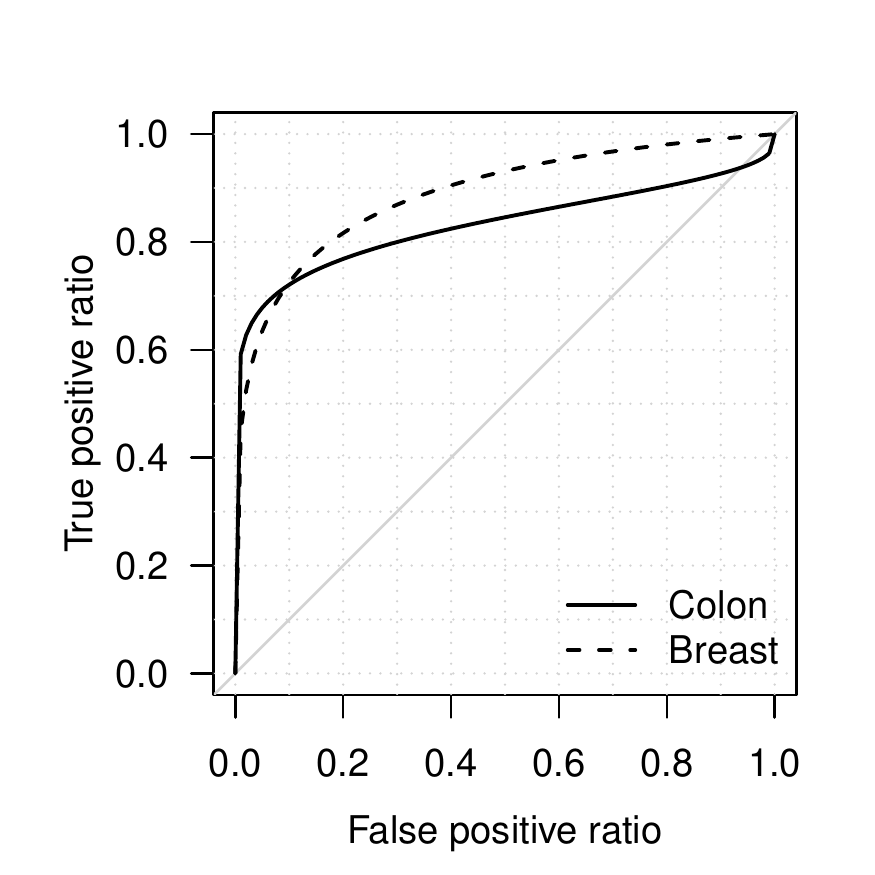}
\includegraphics[width = .435\textwidth, trim = 0 15 0 0, clip]{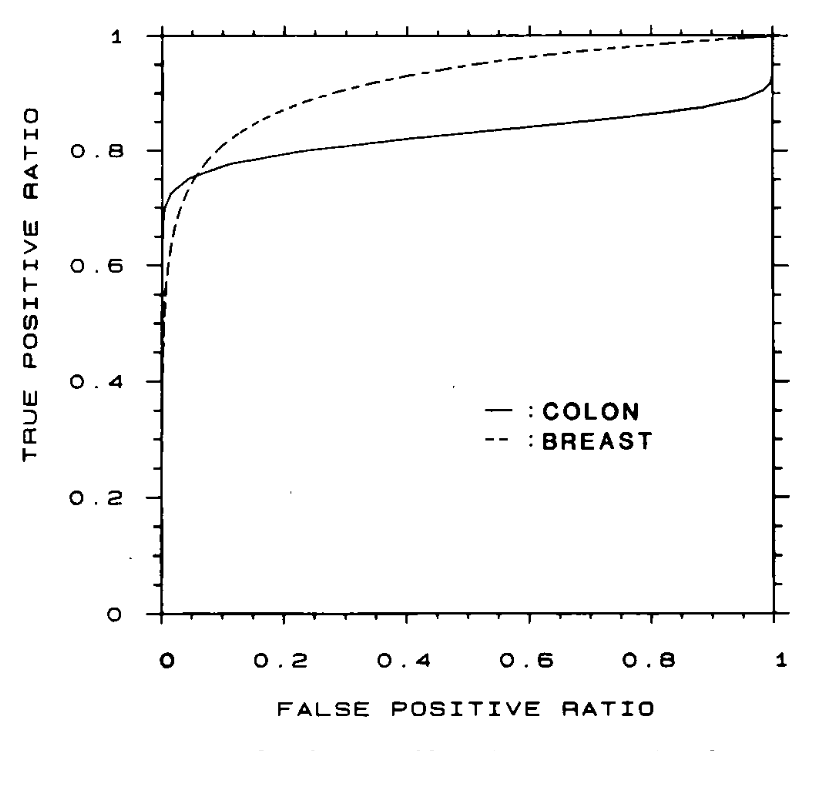}
\caption{Re-analysis of \protect{\cite{Tosteson_1988}}. ROC curves for
ultrasound by primary tumor site estimated by a \ls transformation model (left)
and the original Figure~9 of \protect{\cite{Tosteson_1988}} (right).
\label{fig:ROC}}
\end{figure}

Using the implementation of location-scale transformation models in the
\pkg{tram} \proglang{R}~add-on package \citep{pkg:tram}, we, of course, can
estimate model~(\ref{eq:TB}), replicating the corresponding analysis.  The
smooth receiver operating characteristic (ROC) curves obtained in our
re-analysis are shown in Figure~\ref{fig:ROC} (left), they very closely
reproduce the results of \citet[][Figure~9]{Tosteson_1988}.

\section{Best subset selection algorithm} \label{sec:bss}

The \alg{abess} algorithm \citep{Zhu_2020} can be used to perform best subset 
selection in location-scale transformation models 
(Algorithm~\ref{alg:sabess:supp})
\begin{align*}
  \max_{\parm \in \RR^\dimparm, \shiftparm \in \RR^J, \scaleparm \in \RR^J} 
  \sum_{i=1}^N \ell_i\left(\parm, \rx_i^\top\shiftparm,
  \sqrt{\exp\left(\rx_i^\top\scaleparm\right)}^{-1}\right),
  \quad \mbox{subject to } \norm{\mvec \odot \left(\shiftparm^\top,
  \scaleparm^\top\right)^\top}_0 \leq s.
\end{align*}
Here, $\odot$ denotes element-wise multiplication and the support size $s \in \{1, 
\dots, \norm{\mvec}_0\}$ is fixed. Further, $\norm{\vvec}_0 := \lvert\{j: \vvec_j 
\neq 0 \}\rvert$ denotes the $L_0$-norm and the $2J$-vector $\mvec$ encodes 
mandatory covariates by ex-/including parameter $j$ from penalization via 
$\mvec_j = 0$ or $\mvec_j = 1$, respectively. For instance, when penalizing only 
the scale term $m_j = 0$ for $j = 1, \dots, \dim \shiftparm$ and 1 otherwise. Note 
that the parameters of the transformation function $\parm$ always remain unpenalized. 
The algorithm requires choosing an initial active set $\calA^0_s$ for fixed support
size $s$. \citet{Zhu_2020} recommend using the $k$ covariates most highly correlated 
with the response $\rY$. However, empirical correlations are not well-defined for 
censored responses. Package \pkg{tramvs} \citep{pkg:tramvs} circumvents this by 
instead using the $k$ covariates that are most correlated with the (location or scale)
score residuals of a transformation model containing the mandatory covariates only.
Next, \alg{abess} iteratively updates the active set in the ``splicing'' step 
\citep[for details see Algorithm~2 in][]{Zhu_2020}, which is based on both the
improvement in log-likelihood when including an additional covariate and the
sacrifice when removing an already included covariate. When the support size $s$ is 
unknown, $s$ is tuned based on the minimal high-dimensional information 
criterion (SIC),
\begin{align}\label{eq:SIC}
  \SIC(\parm, \shiftparm, \scaleparm) = - \sum_{i=1}^N \ell_i\left(\parm, 
  \rx_i^\top\shiftparm, \sqrt{\exp(\rx_i^\top\scaleparm)}^{-1}\right) + \norm{\mvec 
  \odot \left(\shiftparm^\top, \scaleparm^\top\right)^\top}_0 (\log 2J) (\log\log N).
\end{align}
\begin{algorithm}[!ht]
\scriptsize
\caption{Best subset selection for \ls transformation models with
mandatory covariates.}\label{alg:sabess:supp}
\begin{algorithmic}[1]
\Require Data $\{(\ry_i, \rx_i)\}_{i=1}^N \in (\Xi \times \RR^J)^N$, 
    max. support size $s_\text{max} \in \{1, \dots, 2J\}$, 
    max. splicing size $k_\text{max} \leq s_\text{max}$, tuning threshold 
    $\tau_s \in \RR_+$ for $s = 1, \dots, s_\text{max}$, vector of mandatory 
    covariates $\mvec \in \{0, 1\}^{2J}$
\State $\calM \gets \{j : \mvec_j = 0\}$ \gComment{Find mandatory covariates}
\State $\calJ \gets \{1 \leq j \leq J : \mvec_j = 1\}$ \gComment{Location covariates for selection}
\State $\calK \gets \{J + 1 < k \leq 2J : \mvec_{j} = 1\}$ \gComment{Scale covariates for selection}
\State $(\hat\parm, \hat\shiftparm, \hat\scaleparm) \gets
    \argmax_{\parm \in \RR^\dimparm, \shiftparm \in \RR^J, \scaleparm \in \RR^J} \sum_{i=1}^N 
        \ell_i\left(\parm, \rx_i^\top\shiftparm, \sqrt{\exp(\rx_i^\top\scaleparm)}^{-1}\right)$
        \newline
        subject to $\operatorname{supp}[(\shiftparm^\top, \scaleparm^\top)^\top] = \calM$
    \gComment{Fit model with mandatory covariates}
\State $(r_\text{loc}, r_\text{sc})_i \gets \left.
    \frac{\partial}{\partial(\eshiftparm,\escaleparm)} 
        \ell_i\left(\parm,  \eshiftparm_0 + \rx_i^\top\hat\shiftparm, 
        \sqrt{\exp(\escaleparm_0 + \rx_i^\top\hat\scaleparm)}^{-1}\right)\right\rvert_{\parm=\hat\parm,\eshiftparm_0=0, \escaleparm_0=0}$
    \gComment{Compute bivariate score residuals}
\For{$s = 1, 2, \dots, s_\text{max}$}
\State Initialize active set \gComment{Find $s$ covariates most correlated with score residuals}
\begin{align*}
\calA^0_s &= \bigg\{i : 
\sum_{j \in \calJ}
    \1(
        \lvert\cor(\rx_j, \rvec_\text{loc})\rvert \geq \lvert\cor(\rx_i, \rvec_\text{loc})\rvert 
    ) \; +
\sum_{k \in \calK}
    \1(
        \lvert\cor(\rx_k, \rvec_\text{sc})\rvert \geq \lvert\cor(\rx_i, \rvec_\text{sc})\rvert 
    )
    \leq s \bigg\}, \\ \calI^0_s &= (\calJ \cup \calK) \backslash \calA^0_s
\end{align*} 
    \For{$m = 0, 1, 2, \dots$} \gComment{See Algorithm~2 in \citet{Zhu_2020}}
        \State $(\hat\parm^{m+1}_s, \hat\shiftparm^{m + 1}_s, \hat\scaleparm^{m+1}_s, \calA_s^{m+1}, \calI_s^{m+1})
        \gets \operatorname{Splicing}(\hat\parm^m_s, \hat\shiftparm^m_s, \hat\scaleparm^m_s, \calA^m_s, 
        \calI^m_s, k_\text{max}, \tau_s)$ 
    \If{$(\calA^{m+1}_s, \calI^{m+1}_s) = (\calA^m_s, \calI^m_s)$} 
    \State stop
    \EndIf
    \EndFor
\State $(\hat\parm_s, \hat\shiftparm_s, \hat\scaleparm_s, \hat\calA_s) \gets 
    (\hat\parm^{m+1}_s, \hat\shiftparm^{m+1}_s, \hat\scaleparm^{m+1}_s, \calA^{m+1}_s)$
\EndFor
\State $\hat s = \argmin_{s} \SIC(\hat\parm_s, \hat\shiftparm_s, \hat\scaleparm_s)$
\gComment{Choose optimal support size based on SIC (Equation~\ref{eq:SIC})}
\State \Return{$(\hat\parm_{\hat s}, \hat\shiftparm_{\hat s}, \hat\scaleparm_{\hat s}, \hat \calA_{\hat s})$}
\end{algorithmic}
\end{algorithm}

\section{Simulation} \label{sec:SIM}

The algorithms employed for estimating stratified models
(Section~\ref{sec:STRAT}), crossing hazards models (Section~\ref{sec:XH}),
partial proportional hazards models (Section~\ref{sec:PPH}), \ls
transformation trees (Section~\ref{sec:TRTF}), and transformation additive
models for location and scale (Section~\ref{sec:TAMLS}) have all been
demonstrated to work well in a plethora of publications.  The best subset
selection algorithm applied to linear \ls transformation models is novel and
thus a more thorough investigation of its empirical performance was called
for.

Data were generated from a linear \ls transformation model 
\begin{eqnarray}
\Prob(\rY \le \ry \mid \rX = \rx) & = & \logit^{-1}\left(\sqrt{\exp(\escaleparm x_2)}
\h(\ry) - \eshiftparm x_1\right), \quad x_1, x_2 \in [0, 1], \label{eq:dgp}
\end{eqnarray}
with $X_j \sim \UD[0,1], j = 1,2$ and corresponding baseline distribution
\begin{eqnarray*}
\Prob(\rY \le \ry \mid \rX = (0, 0)) & = & \pZ_{\chi^2_3}(\ry) \iff
\h(\ry) = \logit(\pZ_{\chi^2_3}(\ry)).
\end{eqnarray*}
Sets of 500~observations were generated from
model~(\ref{eq:dgp}) for each combination of the parameter values
$\eshiftparm \in (0, 0.5, 1)$ and $\escaleparm \in (0, 0.5, 1)$. 
The simulation experiment was repeated 100~times for all parameter configurations.

\begin{figure}[t!]
\begin{knitrout}\small
\definecolor{shadecolor}{rgb}{1, 1, 1}\color{fgcolor}

{\centering \includegraphics[width=.9\textwidth]{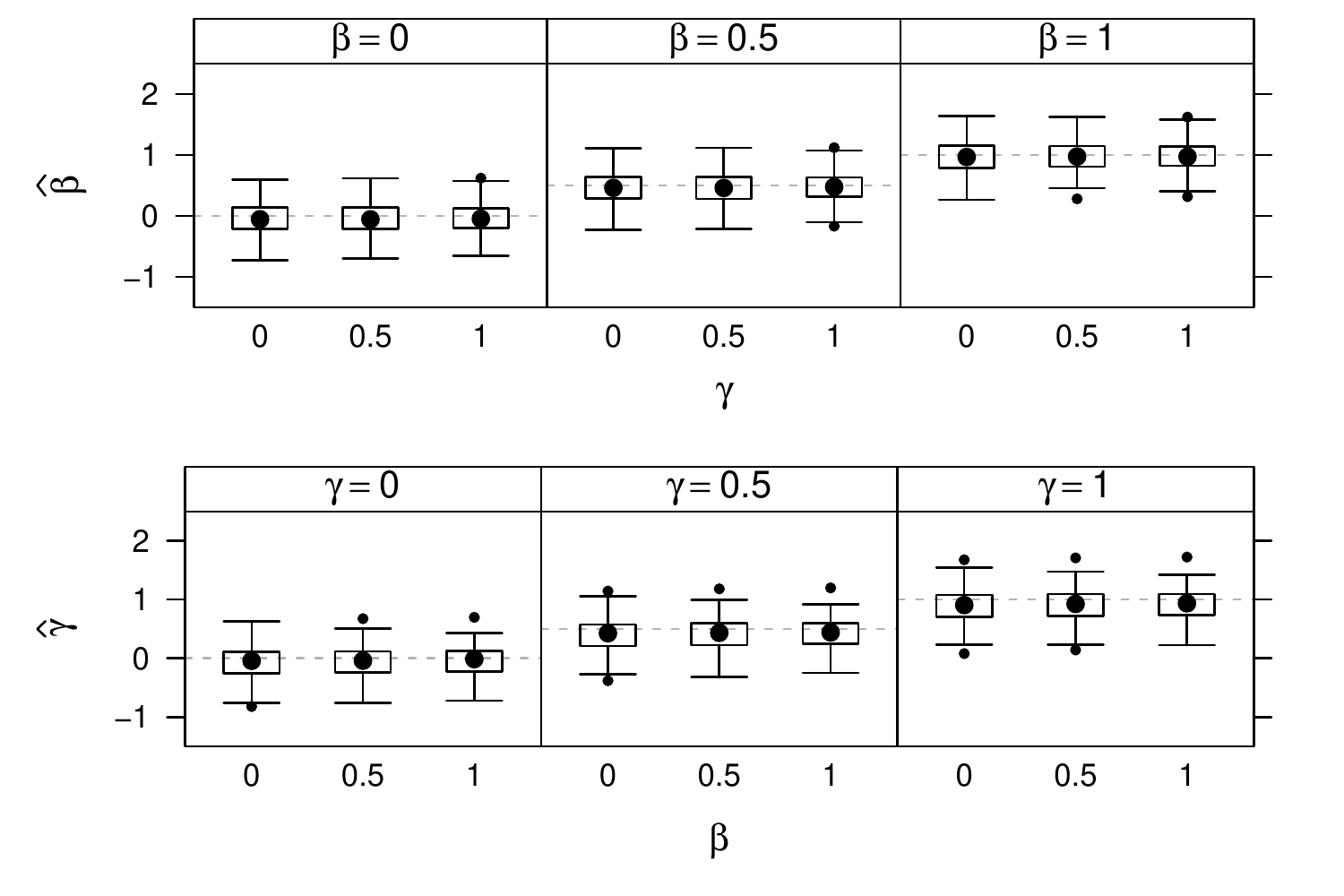} 

}

\end{knitrout}
\vspace*{-3mm}
\caption{Simulation results. \L parameter estimates $\hat\eshiftparm$ (first
row) and scale parameter estimates $\hat\escaleparm$ (second row) employing full
maximum likelihood estimation are shown along all configurations of $\eshiftparm$ and $\escaleparm$.
The true value of the corresponding parameter is indicated by the dashed horizontal line.
\label{fig:SIM:ML}}
\end{figure}

As a benchmark method, full maximum likelihood estimation of the model
parameters was chosen.  The results in Figure~\ref{fig:SIM:ML},
comparing the \l and scale parameter
estimates, $\hat\eshiftparm$ and $\hat\escaleparm$,
to the true parameter values, corroborate that both model parameters
can be precisely estimated, with the corresponding variance staying
constant along all magnitudes of the true parameter values.

\begin{figure}
\begin{knitrout}\small
\definecolor{shadecolor}{rgb}{1, 1, 1}\color{fgcolor}

{\centering \includegraphics[width=.9\textwidth]{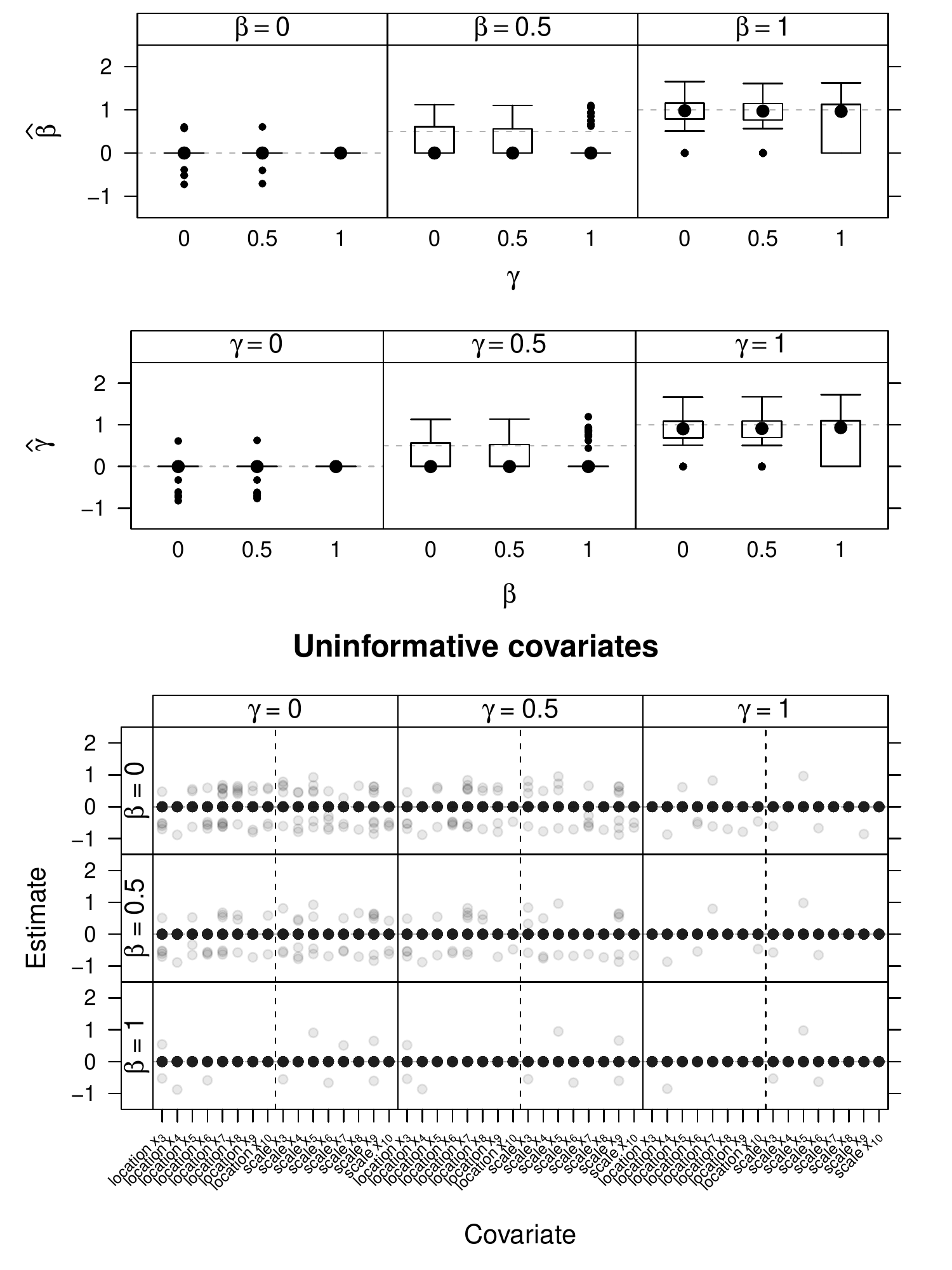} 

}

\end{knitrout}
\vspace*{-3mm}
\caption{Simulation. \L parameter estimates $\hat\eshiftparm$ (first
row) and scale parameter estimates $\hat\escaleparm$ (second row) employing the
subset selection algorithm are shown along all configurations of $\eshiftparm$ and $\escaleparm$ (top).
The estimates corresponding to the uninformative covariates employing the best 
subset selection algorithm are shown in the bottom.
The true value of the corresponding parameter is indicated by the dashed horizontal line.
\label{fig:SIM:VS}}
\end{figure}

To contrast the best subset selection algorithm, additional variables $X_j
\sim \UD[0, 1],$ $j=3, \dots, 10$ with $\eshiftparm_j = \escaleparm_j = 0$
were added to the \l and scale term of
model~(\ref{eq:dgp}). 
The corresponding model parameters, estimated using the best subset selection
algorithm, are shown in Figure~\ref{fig:SIM:VS}.
Similar to the benchmark method, the parameter estimates, $\hat\eshiftparm$ and
$\hat\escaleparm$, employing the best subset selection algorithm,
indicate that the parameters can be well estimated, however with some 
distortion towards zero (Figure~\ref{fig:SIM:VS}, top).
Assessing the estimates corresponding to the
uninformative covariates ``\l $\erx_j$'' and ``scale $\erx_j$'' (with true
value $\eshiftparm_j = \escaleparm_j = 0$), reveal that in the best
subset model the uninformative variables are mostly correctly negated, especially for
configurations with larger values of $\eshiftparm$ or $\escaleparm$
(Figure~\ref{fig:SIM:VS}, bottom).
 \end{appendix}
\clearpage 
\end{document}